\documentclass[prd,floats,preprint,superscriptaddress,floatfix,eqsecnum,nofootinbib,12pt]{revtex4}

\pdfoutput=1

\usepackage{amsmath,amssymb,graphicx,bigstrut,ulem,tensor,hyperref,xcolor}

\def\be{\begin{equation}}
\def\ee{\end{equation}}
\def\beq{\begin{eqnarray}}

\def\eeq{\end{eqnarray}}
\def\Dobs{D_{\rm uloc}}
\def\Dobsaoi{D^{\ge 1}_{\rm uloc}}
\def\D1{D^{\ge 1}}
\def\cO{{\cal O}}
\def\Dell{D_{\rm loc}}
\def\bigstar{\mbox{\Large $*$}}
\def\bigdot{\mbox{\Large $\bullet$}}
\def\FT{(I,\Psi)}

\begin{document}

\title{Sharp Predictions from Eternal Inflation Patches \\ in D-brane Inflation}

\author{Thomas Hertog}
\email{thomas.hertog@fys.kuleuven.be}
\affiliation{Institute for Theoretical Physics, KU Leuven, 3001 Leuven, Belgium}
\author{Oliver Janssen}
\email{opj202@nyu.edu}
\affiliation{Center for Cosmology and Particle Physics, NYU, NY 10003, USA}

\pagenumbering{roman}

\begin{abstract}

We numerically generate the six-dimensional landscape of D3-brane inflation and identify patches of eternal inflation near sufficiently flat inflection points of the potential. We show that reasonable measures that select patches of eternal inflation in the landscape yield sharp predictions for the spectral properties of primordial perturbations on observable scales. These include a scalar tilt of .936, a running of the scalar tilt $-$.00103, undetectably small tensors and non-Gaussianity, and no observable spatial curvature. Our results explicitly demonstrate that precision cosmology probes the combination of the statistical properties of the string landscape and the measure implied by the universe's quantum state.

\end{abstract}

\pacs{98.8.Qc, 98.8.Bp, 98.8.Cq, 04.6.-m}

\maketitle

\tableofcontents

\newpage

\section{Introduction}
\label{intro}
\setcounter{page}{0}
\pagenumbering{arabic}

A framework for cosmology consisting of a model of the dynamics together with a theory of initial conditions {\it predicts} a prior. The prior takes the form of probability distributions for cosmological observables. Observables with distributions that are sharply peaked around specific values are predicted with high accuracy by the theory and can be used to test it against observations.

In this paper we compute the probabilities for CMB-related observables in a specific corner of the string landscape, the six-dimensional landscape of warped D3-brane inflation, for a universe in the no-boundary quantum state. We will find that a number of observables associated with the CMB which have rather broad statistical distributions in the landscape are nevertheless sharply predicted when the implications of the universe's quantum state are taken into account. Although we mostly focus on a particular quantum state, we will argue that many of the sharp predictions persist in other viable theories of initial conditions that select among all quantum mechanical histories of the universe those with a period of slow roll inflation.

We construct the D3-brane inflationary landscape by modeling the six-field dynamics of D3-brane inflation on a general scalar potential on the conifold. The scalar dynamics is governed by a potential which is fixed up to a set of compactification-dependent constants. To realise this landscape we choose values for these constants with a Monte Carlo method \cite{Agarwal2011,Dias2012,Mcallister2012}. Previous realisations have identified patches near approximate inflection points of the potential which allow for an extended period of slow roll inflation. However the ensemble of inflationary universes in the landscape exhibits a broad range of values for observables such as the scalar spectral tilt $n_s$, the running of the tilt $\alpha_s$ and the tensor-to-scalar ratio $r$ of the fluctuations. For instance in universes with less than 60 efolds of inflation after the inflection point the scalar spectrum is typically blue, whereas otherwise it is red. Moreover the landscape does not appear to have a strong statistical bias toward either option\footnote{In the subset of trajectories with $N_e > 60$ there is a moderate statistical bias towards a blue spectrum \cite{Dias2012}. As we will show, this bias is not relevant for predictions.}. This has led to broadly distributed probabilities - and hence no clean predictions - for CMB-related observables if one adopts a flat prior on initial conditions reflecting ignorance about the underlying theory of the quantum state.

By contrast, combining the statistics of the landscape with a model of the quantum state yields a very different conclusion. In this paper we mostly adopt the semiclassical no-boundary wave function (NBWF) as a model quantum state \cite{Hartle1983}. The NBWF populates a given landscape potential with an ensemble of universes -- backgrounds {\it and} fluctuations -- and, most importantly, it endows this ensemble with a measure. At the semiclassical level the NBWF weights universes by $\exp(-I)$ where $I$ is the Euclidean action of a compact regular saddle point that smoothly joins onto the universe when the scale factor is sufficiently large\footnote{We use units where $\hbar = c = 8\pi G = 1$.}.

It is well known that the probability distribution over cosmological backgrounds predicted by the NBWF is concentrated around inflationary universes (see e.g. \cite{Hartle2007,Hartle2008}). Hence the NBWF {\it selects} those patches of the landscape where the slow roll conditions for inflation hold. The sum of the probabilities of all universes originating in a given inflationary patch of the landscape implies a relative weighting of the `models' of inflation contained in the landscape, which then yields a prior over various observables \cite{Hartle2010b,Hertog2013}.

We shall be primarily interested in the joint probabilities of observables connected to the statistical features of CMB perturbations, such as the scalar tilt $n_s$, the tensor-to-scalar ratio $r$ and their overall amplitude $A_s$. The NBWF predicts the usual probabilities for nearly Gaussian scalar and tensor perturbations around each inflationary background in its ensemble \cite{Halliwell1985,Hartle2010b}. Their statistical properties are specified by the shape of the potential patch probed by the background. This is where the relative weighting of different landscape regions enters: if the NBWF prior is sharply peaked around inflationary backgrounds associated with a particular patch of the landscape then the theory predicts the CMB perturbation spectrum should exhibit the features characteristic of the potential in that patch.

However probabilities that are relevant for the prediction of our observations are not simply the probabilities for members of the ensemble of histories of the whole universe. Rather they are conditional probabilities that describe correlations between local cosmological observables ${\cal O}$, given our observational situation. We have called the former bottom-up (BU) probabilities and the latter top-down (TD) probabilities \cite{Hawking2006}. The condition that our observational situation exists somewhere in the universe is a global condition that selects histories in which our classical universe emerges from a region in the landscape where the conditions for eternal inflation (EI) hold \cite{Hartle2009,Hartle2010}. A region of EI is a very flat patch of the landscape, associated with scales beyond those directly probed by CMB fluctuations, where $V>\epsilon$, with $\epsilon \equiv V'^2/2V^2$. The quantum dynamics of perturbations implies that universes originating in such regions become exceedingly large, which in turn means they dominate TD probabilities in a broad class of quantum states - provided of course there are EI patches.

It was argued \cite{Chen2006} that the D-brane inflationary landscape has no (slow roll) EI patches, at least when one also requires a low amplitude of scalar fluctuations on observable scales. However the analysis of \cite{Chen2006} did not incorporate flux-induced contributions to the potential (because they were not understood at the time) and used an EI condition of the form $\phi > \phi_c$, which does not allow for the possibility that subtle cancellations between terms in the potential could lead to sufficiently flat patches where EI might occur. A new investigation of the existence of EI patches in the D-brane landscape is thus called for. With the addition of the missing terms (and dimensions) and by a Monte Carlo scan of the potential landscape, we find that slow roll eternal inflation is statistically rare, but nonetheless possible.

The TD probabilities predicted by the NBWF are biased towards universes associated with regions of EI in the landscape at a low value of the potential \cite{Hartle2010,Hertog2013}. We find that in the D3-brane landscape universes emerging from low-lying regions of EI subsequently undergo a long period of slow roll inflation. Moreover the spectral features of primordial fluctuations on observable scales in those histories have very specific statistical features, such as a scalar tilt $n_s \approx .936$ and a tensor-to-scalar ratio $\log_{10} r < -13$. It is immediately evident that these specific and sharp predictions follow from the combination of the NBWF prior with the statistical structure of the landscape. Indeed earlier work has shown that either part of the theory separately leads to much more broadly distributed predictions for these observables, demonstrating that precision observations in cosmology probe an intricate combination of the dynamics and the quantum state of the universe.

The rest of this paper is organised as follows. In Section \ref{landscape} we describe the inflationary landscape of warped D-brane inflation. Section \ref{branereview} gives a brief overview of the model. In Section \ref{approxrestrsec} we state our notation and approximations. Section \ref{landscapesec} contains the explicit form of the landscape potential, and finally in Section \ref{ensembleconstr} we discuss how we generate an ensemble of representative trajectories for D-brane inflation building on  \cite{Agarwal2011,Mcallister2012,Dias2012}. However in contrast with \cite{Agarwal2011,Mcallister2012,Dias2012} we allow a specific model parameter $a_0$, the warp factor at the tip of the conifold, to vary. This enables us to identify patches of eternal inflation, which is the main subject of Section \ref{EIsec}. We first review the subset of inflating trajectories with $N_e > 60$ and no further requirements, which was discussed in \cite{Agarwal2011,Mcallister2012,Dias2012}. In Section \ref{EI} we concentrate on eternal inflation in D-brane inflation and show some of the statistical properties of the subset of EI patches. In Section \ref{NBWF} we turn to quantum cosmology and introduce the notion of a wave function of the universe, which effectively acts as a theory of initial conditions on the landscape. We focus on a specific wave function, the previously mentioned NBWF, in Section \ref{NBWFsec}. In Section \ref{probobs} we review how predictions for observations are derived from a wave function of the universe, and we apply the general framework to the specific case of D-brane inflation. In Section \ref{testing} we specify precisely what are the conditional probabilities involving CMB-related observables we compute, and we discuss the results for these in Section \ref{probabilitysec}. Our findings can be explained in terms of a simple phenomenological model of single-field inflection point inflation, which we work out in Section \ref{inflectionptmodel}. In Section \ref{othermeasures} we briefly consider predictions following from alternative choices of measure or for other observables, and we conclude with a comparison with observations in Section \ref{conclusion}.

\section{Landscape: Warped D-brane Inflation}
\label{landscape}

\subsection{Overview}
\label{branereview}

In warped D-brane inflation\footnote{Our discussion of D-brane inflation is based on \cite{Baumann2010,Agarwal2011,Baumann2014} where additional details may be found. Most of our notation follows \cite{Agarwal2011}.}, one considers flux compactifications of type IIB string theory on Calabi-Yau threefolds where the moduli-stabilising fluxes are chosen such that the internal space exhibits a warped throat. More specifically, one approximates the throat by a finite region of the non-compact warped deformed conifold \cite{Klebanov2000} and assumes that it smoothly attaches onto the bulk Calabi-Yau space at a radial distance $r \approx r_{\tiny \mbox{UV}}$. The background geometry, produced by a stack of $N$ D3-branes at the tip of the singular conifold and $M$ D5-branes wrapped around collapsed 2-cycles in the tip, is described by the string frame metric
\be \label{warpedmetric}
\mathrm{d}s^2 = e^{2A(y)} g_{\mu \nu} \mathrm{d}x^\mu \mathrm{d}x^\nu + e^{-2A(y)} G_{AB}(y) \mathrm{d}y^A \mathrm{d}y^B.
\ee
Here, $G$ is the deformed conifold metric, $e^A$ is a warp factor which we specify in Section \ref{approxrestrsec} and $g$ is the metric in the four large dimensions. The NBWF predicts the four-dimensional background to be a closed FLRW geometry,
\be \label{FLRW}
-\mathrm{d}t^2 + a(t)^2 \left( \frac{\mathrm{d}r^2}{1-\kappa r^2} + r^2 \mathrm{d}\Omega^2 \right),
\ee
with $\kappa >0$. In the KKLMMT scenario \cite{Kachru2003}, which we consider, inflation is driven by the interaction of a mobile D3-brane and a stationary anti-D3-brane which is located at the tip of the cone where it minimises its energy. The branes fill our three large spatial dimensions nearly homogeneously and are pointlike in the six extra dimensions. Inflation ends when (and if ever) the branes come close enough together to annihilate through tachyon condensation.

The brane dynamics in the throat and sufficiently far away from the strongly deformed tip are governed by an effective potential which contains terms of distinct origin: an attractive Coulomb-like term due to the opposite R-R charge of the branes, a term (also attactive) due to the coupling to four-dimensional curvature\footnote{This is just one term to the order in which we work, see Section \ref{landscapesec}.} and bulk terms - corrections to the potential due to gluing the throat onto the bulk Calabi-Yau and the presence of compactification artifacts e.g. flux. A delicate balance between these competing forces on the mobile brane can lead to an extended period of inflation. In general the brane meanders through all six extra dimensions before colliding with the antibrane, making D-brane inflation an example of multifield inflation.

Through the work of \cite{Baumann2010} it has become possible to characterise the flux-induced potential by a countably infinite set of real numbers called Wilson coefficients. Ideally one would be able to fix the Wilson coefficients for a specific situation in which all compactification details are known, but the present state of affairs \cite{Baumann2010} typically does not allow for this - only an order of magnitude estimate of the coefficients can be made \cite{Agarwal2011,Baumann2008}. A way forward was provided in \cite{Agarwal2011} by sampling the Wilson coefficients from some distribution $\mathcal{M}$. Together with a choice of truncation degree of the potential\footnote{The potential can be written as an expansion in increasing powers of a dimensionless quantity $x < 1$, see Section \ref{landscapesec}.}, initial conditions and remaining model parameters, one generated an ensemble of realisations of D-brane inflation, i.e. a collection of brane trajectories. Remarkably, one was able to identify a statistical feature of this ensemble that is substantially independent of $\mathcal{M}$, truncation degree and initial conditions, namely the fraction of trials $N(N_e)$ exhibiting $N_e$ efolds of inflation which was shown to behave like\footnote{For fixed model parameters. We have verified that the independence also holds for a range of values of the important parameter $a_0$.} $N_e^{-3}$. An analytic argument for this characteristic behavior can be made by considering the effectively single-field trajectories, which form the majority of realisations producing enough inflation ($\gg$ 70 efolds), for which the potential has a typical inflection point \cite{Agarwal2011,Mcallister2012}. This emergent universality, which the authors of \cite{Agarwal2011} ascribe to the collective structure that is likely to arise in a potential with many competing terms, is a reason to believe that the statistical distribution of cosmological observables in the actual corner of the string landscape defined by D-brane inflation can be probed by making definite (though not pathological) choices for $\mathcal{M}$, truncation degree of the potential, initial conditions and model parameters which together define a toy model of the landscape. This reasoning was adopted in subsequent statistical analyses of cosmological perturbations in D-brane inflation \cite{Mcallister2012,Dias2012} and is a basic assumption of our work. If future work would show that $\mathcal{M}$ is in fact a very specific distribution, perhaps introducing significant correlations between the Wilson coefficients, which would either prohibit EI or change the phenomenology away from inflection point inflation, our predictions for D-brane inflation must be reconsidered.

\subsection{Approximations and Restrictions}
\label{approxrestrsec}

In our analysis, following \cite{Baumann2010,Agarwal2011,Mcallister2012,Dias2012}, we restrict ourselves to the regime well above the tip of the cone, where we may neglect the deformation and approximate the deformed conifold metric $G$ in \eqref{warpedmetric} by that of the singular conifold\footnote{For a discussion of the conifold in all its incarnations, see \cite{Candelas1989,Gwyn2007}.}
\be \label{metricconifold}
G_{AB}(y) \mathrm{d}y^A \mathrm{d}y^B = \mathrm{d}r^2 + r^2 \mathrm{d}\Omega_{T^{1,1}}^2,
\ee
where $\mathrm{d}\Omega_{T^{1,1}}^2$ is the Sasaki-Einstein metric on the base $T^{1,1} = \big( \mbox{SU(2)} \times \mbox{SU(2)} \big) / \mbox{U(1)}$ of the conifold,
\be \label{T11metric}
\mathrm{d}\Omega_{T^{1,1}}^2 = \frac{1}{6}\displaystyle\sum_{i=1}^2 \left( \mathrm{d}\theta_i^2 + \sin^2 \theta_i \mathrm{d}\phi_i^2 \right) + \frac{1}{9} \left( \mathrm{d}\psi + \displaystyle \sum_{i=1}^2 \cos\theta_i \mathrm{d}\phi_i \right)^2.
\ee
So we describe the conifold by six real coordinates, one radial coordinate $r$ and five angular coordinates $\theta_{1,2} \in [0,\pi]$, $\phi_{1,2} \in [0,2\pi]$ and $\psi \in [0,4\pi]$ which we have abbreviated by $y = (r,\theta_1,\theta_2,\phi_1,\phi_2,\psi) \equiv (r,\Psi)$. In our analysis of the brane dynamics, we write $\phi^A$ for the $A$-th coordinate of the mobile brane in the sequence $(r,\Psi)$ which appears as a scalar field living on the conifold in the four-dimensional effective Lagrangian and action given below in \eqref{DBI} and \eqref{effectiveaction}. \\
In our regime of study, the warp factor in $\eqref{warpedmetric}$ may be well-approximated by $e^A = r/R$ \cite{Baumann2014} where $R^4 = \frac{27 \pi}{4} g_s N (\alpha')^2$. The radial distance $r_{\tiny \mbox{UV}}$, where the conifold glues onto the bulk, is approximately equal to $R$. The deformation becomes significant in the vicinity of $r \approx a_0 R$, where as mentioned above $a_0 = e^{A_{\tiny \mbox{IR}}}$ is the warp factor at the tip of the cone. Concretely, in terms of a rescaled radial coordinate $x \equiv r / r_{\tiny \mbox{UV}}$, we restrict our analysis to the region $10 a_0 < x < 1$ of the cone\footnote{This is less restrictive than \cite{Agarwal2011,Mcallister2012,Dias2012} but we do not believe it to be problematic. The D-brane potential is currently not known near the tip of the warped deformed conifold, but recently has been computed for warped \textit{resolved} conifold compactifications in \cite{Kenton2014}. In this context there is the outstanding issue of stabilising the closed string moduli, however.}.
The restriction to stay away from the tip reduces the likelihood of DBI effects, see Section \ref{ensembleconstr} and \cite{Agarwal2011}. Finally, when discussing perturbations of the homogeneous background, one must show that all entropic modes have decayed by the time the brane leaves the restricted volume of the conifold for predictions to be valid, i.e. independent of further evolution in the tip-regime and the reheating process. We have verified that the entropic modes are always sufficiently massive such that we would expect them to decay quickly after Hubble exit (see also Section \ref{EI}), but we rely on previous studies \cite{Mcallister2012,Dias2012} that assure an adiabatic limit is always reached.

\subsection{The Potential Landscape}
\label{landscapesec}

It is the achievement of \cite{Baumann2010} to determine the most general form of the potential $V$ governing the brane dynamics. One shows that the rescaled potential $V/T_3 = \Phi_{-}$ satisfies the Poisson equation
\be \label{potentialeq}
\nabla^2 \Phi_{-} = \frac{g_s}{96} |\Lambda|^2 + R_4 + \mathcal{S}_{\tiny \overline{\mbox{D3}}},
\ee
where $\nabla^2$ is the Laplacian on the singular conifold, $T_3 = \big( (2 \pi)^3 g_s (\alpha')^2 \big)^{-1}$ is the D3-brane tension and one recognises the distinct sources we mentioned in Section \ref{branereview}. It is shown in \cite{Baumann2010} that the perturbing imaginary anti-self-dual (IASD) three-form flux $\Lambda$ on a Calabi-Yau cone, in particular on the singular conifold, can be categorised into three groups. Each type of flux is expressed as a function of a harmonic function on the conifold, its derivatives and objects related to the K\"ahler structure of the conifold - the K\"ahler potential, K\"ahler form and holomorphic 3-form. A general harmonic function $f$ on the conifold may be expressed in a multipole expansion of the form
\be \label{harmonic}
f(x,\Psi) = \sum_{L,M} c_{\tiny{LM}} x^{\Delta(L)} Y_{LM}(\Psi) + \mbox{c.c. },
\ee
where $L \equiv (l_1, l_2, R)$ and $M \equiv (m_1, m_2)$ label the SU(2) $\times$ SU(2) $\times$ U(1) quantum numbers under the isometries of $T^{1,1}$, which are restricted by group-theoretic selection rules \cite{Ceresole1999} (see below \eqref{Vfluxrealdeal} for the first few allowed values). The $Y_{LM}$ are the angular harmonics on $T^{1,1}$ with eigenvalues $-H(L)$, i.e. $\nabla^2_\Psi Y_{LM} = -H(L) Y_{LM}$, where 
\be \label{eigenvalues}
H(l_1,l_2,R) = 6 \left( l_1(l_1+1) + l_2(l_2 + 1) - \frac{R^2}{8} \right),
\ee
which can be found in explicit form in e.g. \cite{Klebanov2007}. Demanding $\nabla^2 f = 0$ in \eqref{harmonic} then implies that the radial scaling dimensions $\Delta$ are related to $H$ via
\be \label{radialscaling}
\Delta = -2 + \sqrt{H + 4}.
\ee
Finally, the constants $c_{LM}$ are the undetermined Wilson coefficients we mentioned earlier. The general solution to \eqref{potentialeq} is then the sum of a solution to the homogeneous equation of the form \eqref{harmonic}, plus inhomogeneous terms which can be obtained by integrating \eqref{potentialeq} over the Green's function on the singular conifold. The interested reader is redirected to \cite{Baumann2010} where one explains the solution in detail. Here we simply display the final result in analogy with \cite{Agarwal2011} but in slightly more detail:
\be \label{potentialschematic}
V(x,\Psi) = V_{C}(x) + V_{R}(x) + V_{\tiny \mbox{hom}}(x,\Psi) + V_{\tiny \mbox{flux}}(x,\Psi),
\ee
with
\be \label{coulombpot}
V_{C}(x) = D_0 \left( 1 - \frac{27 D_0}{64 \pi^2 T_3^2 r_{\tiny \mbox{UV}}^4} \frac{1}{x^4} \right),
\ee
\be
V_{R}(x) = \frac{1}{3} \mu^4 x^2,
\ee
\be \label{Vhomrealdeal}
V_{\tiny \mbox{hom}}(x,\Psi) = \mu^4 \sum_{L,M} c_{\tiny{LM}} x^{\Delta(L)} Y_{LM}(\Psi) + \mbox{c.c. },
\ee
\be \label{Vfluxrealdeal}
V_{\tiny \mbox{flux}}(x,\Psi) = \mu^4 \sum_{L',M'} c_{L'M'} \hspace{1mm} x \hspace{1mm} Y_{L'M'}(\Psi) + \mu^4 c_2 x^2 + \mbox{c.c. },
\ee
where $L \in \{(1/2,1/2,1)_{\Delta = 3/2}, (1,0,0)_{\Delta = 2}, (0,1,0)_{\Delta = 2}\}$, $M \in \{(-l_1...l_1, -l_2...l_2)\}$, $L' \in \{(0,0,0), (1,0,0), (0,1,0), (1,1,0)\}$, $M' \in \{(-l'_1...l'_1, -l'_2...l'_2)\}$, $D_0 \equiv 2 T_3 a_0^4$ and $\mu^4 \equiv r_{\tiny \mbox{UV}}^2 T_3 D_0$.

Several remarks are in order. Firstly, we have assumed that contributions to the inflationary vacuum energy $V_0$ other than the contribution from the brane-antibrane pair $D_0$ are much smaller in magnitude than $D_0$. Thus $D_0 \sim a_0^4$ sets the scale of inflation (we will fix $T_3$). In particular we do not follow \cite{Mcallister2012} where one assumes that $V_0$ is adjusted such that the scalar amplitude takes on its observed value. In Sections \ref{probabilitysec} and \ref{conclusion} we discuss how this assumption plausibly has no effect on our predictions. Note that in addition to setting the scale of inflation, $a_0$ determines the relative strength of the Coulomb and `random' forces on the brane. Secondly, a constant $\mu^4$ has been extracted in \eqref{Vhomrealdeal} and \eqref{Vfluxrealdeal} so that the $c_{LM}$ are expected to be of order 1 in a typical compactification \cite{Baumann2008}. Following \cite{Agarwal2011}, we assume all possible terms are present and typically of the same order of magnitude in the landscape. Thirdly, we have neglected extra terms in $V_C$ which are suppressed by additional powers of $a_0$, which is small, see Section \ref{ensembleconstr}. Fourthly, we have truncated the potential at $\Delta = 2$. Truncating the potential at this arguably minimal exponent - the leading contribution from curvature goes like $x^2$ - is a luxury we permit ourselves due to the universality results of previous work\footnote{On the other hand we note that including additional terms could lead to the discovery of more EI realisations, but it will not likely affect predictions - see Section \ref{EI}.} \cite{Agarwal2011}. There are 27 Wilson coefficients in our truncated potential.

\subsection{Generating an Ensemble of Trajectories}
\label{ensembleconstr}

The equations of motion of the mobile brane in the Einstein frame are obtained from the DBI + CS Lagrangian
\be \label{DBI}
\mathcal{L} = a^3 \left( -T(\phi) \sqrt{1 + \frac{T_3 G_{AB} \partial_\mu \phi^A \partial^\mu \phi^B}{T(\phi)}} - V(\phi) + T(\phi) \right),
\ee
where $T(\phi) = T_3 e^{4A} = T_3 x^4$ is the warped tension and $V$ is the effective D3-$\overline{\mbox{D3}}$-brane potential \eqref{potentialschematic}. As in previous studies \cite{Agarwal2011,Dias2012}, we found that it was consistent to neglect DBI effects\footnote{In all simulations of the dynamics, the quantity $\left(T_3 G_{AB}\dot{\phi}^A \dot{\phi}^B\right) / T(\phi)$ never exceeded $10^{-9}$.} and approximate the effective action of the total system by that of six scalar fields living in the singular conifold, coupled to 4D gravity like
\be \label{effectiveaction}
S = \int \mathrm{d}^4 x \sqrt{- \det g} \left( \frac{1}{2}R_4 - \frac{1}{2} T_3 G_{AB} \partial_\mu \phi^A \partial^\mu \phi^B - V \right).
\ee
In the following we redefine $\phi^1 \rightarrow \sqrt{T_3} \phi^1$, but we will continue to use $x \in [10a_0,1]$ as the radial position of the brane. The equations of motion for the brane coordinates and homogeneous background geometry that follow from this action are
\begin{align}
&\mathcal{D}_t \dot{\phi}^A + 3H\dot{\phi}^A + G^{AB} \partial_B V = 0 \hspace{1.5cm} A = 1...6, \label{scalareqns} \\
&3H^2 = \frac{1}{2} | \dot{\phi} |^2 + V - 3 \frac{\kappa}{a^2}, \label{friedmannagain} \\
&\dot{H} = - \frac{1}{2} | \dot{\phi} |^2 + \frac{\kappa}{a^2}, \label{hdoteq}
\end{align}
where $\mathcal{D}_t \xi^A \equiv \dot{\xi}^A + \Gamma^A_{BC} \dot{\phi}^B \xi^C$, $\partial_B \equiv \partial / \partial \phi^B$ and by the magnitude $|\xi|$ of a vector $(\xi^A)$ we mean $\sqrt{G_{IJ}\xi^I \xi^J}$. \\

To generate an ensemble of trajectories we largely follow the procedure of \cite{Agarwal2011,Dias2012}. As mentioned in Section \ref{branereview}, a realisation of D-brane inflation is determined by a choice of $\mathcal{M}$ from which to sample the Wilson coefficients, truncation degree of the potential, initial conditions and remaining model parameters. We choose $\mathcal{M}$ to be Gaussian with mean zero and standard deviation $Q \in [0; .3]$. This choice of $Q$-range is consistent with the scaling arguments presented in Appendix A of \cite{Baumann2008}, and the upper bound is motivated by the result that we found no eternally inflating realisations - which dominate predictions in the no-boundary state as we argue in Section \ref{probobs} - for $Q > .3$ in a preliminary scan with $Q \in [0,1]$. Because eternal inflation patches become very rare within the ensemble of inflating trajectories when we demand a small scalar amplitude of perturbations, to optimise numerical efficiency we restricted the parameter space as much as possible without loss of generality. When scanning the landscape, we sample $Q$ from the interval $[0;.3]$ with uniform measure. We will argue in Section \ref{probabilitysec} that our predictions are largely independent of this choice. \\

For the truncation of the potential, we neglect terms $\mathcal{O}(x^{p>2})$ as we discussed in the previous section. As initial conditions for the brane, we choose $x(0) = .9$, $\Psi(0) = 1$ and $\dot{x}(0) = 0$, $\dot{\Psi}(0) = 0$, where 0 denotes the time at the start of inflation where the Euclidean NBWF saddle point matches on to the Lorentzian evolution of the universe. At this time, $a(0) \approx 1/V(\phi_0)$ which we rescale to 1, which sets $\kappa = V(\phi_0)^2$ in the closed FLRW \eqref{FLRW}. The (in)dependence of results on initial conditions has been discussed in \cite{Agarwal2011}. For the remaining parameters, following \cite{Agarwal2011,Dias2012}, we set $r_{\tiny \mbox{UV}} = 1$ and $T_3 = 10^{-2}$. For $a_0$ we take a different approach. The statistical distribution of $a_0$ in the landscape is unknown - it is determined by the quantisation of the background three-form fluxes in the warped throat and the total compactification volume through $g_s$ \cite{Baumann2014} - but it is assumed that string theory allows it to vary over a set of stable compactifications. Therefore we will treat it, like the Wilson coefficients, as another landscape parameter which may vary. Specifically, we scan over $\log_{10} a_0 \in [-5,-2]$ with uniform measure at the beginning of each trial. The upper bound $a_0 = 10^{-2}$ is motivated by the consistency of the set-up: recall that the singular conifold approximation is only valid roughly in the regime $x > 10a_0$. So for these realisations we cannot reliably model the dynamics below $r < .1$ which already may seem too substantial to trust any results. However, in succesful realisations of inflation (see Section \ref{inflation}), the radial height of the inflection point around which the inflaton is typically in slow roll is strongly positively correlated with $a_0$. This is to be expected from the potential \eqref{potentialschematic} - \eqref{Vfluxrealdeal}: if $a_0$ is greater, the Coulomb force is stronger and the inflaton will find the balance between the competing forces higher up in the cone. We verified that in all our realisations, the brane has exited the slow roll phase and has begun its rapid descent towards the tip before our approximations become invalid. Unless the tip-regime contains another slow roll patch or somehow prohibits the branes from annihilating, which is a caveat for all of our realisations, we believe our approximations are justified. Of course the precise cutoff at $\log_{10} a_0 = -2$ is arbitrary, but the conditional probabilities for the observables we consider in this paper are independent of the precise value (see Sections \ref{probabilitysec} and \ref{othermeasures}). The lower bound $a_0 = 10^{-5}$ is motivated by the observation that the amplitude of scalar perturbations in all realisations with $\log_{10} a_0 \lesssim -5$ is smaller than the observed value. We show in Section \ref{probabilitysec} this means these realisations do not contribute to the TD probabilities for the observables we consider. In principle we do not have the liberty to choose to scan logarithmically over $a_0$: ideally, as was the case with $Q$, the statistical distribution of $a_0$ in the D-brane landscape should be dictated by the theory of flux compactifications, which alas is not explicit enough about this matter to date. We will argue however (Section \ref{probabilitysec}), that, like $Q$, predictions following from the no-boundary prior are very robust with respect to the actual distribution of $a_0$ in the landscape. In particular, a uniform or exponential statistical distribution of $a_0$ gives the same predictions for the D-brane landscape in the no-boundary state\footnote{From a practical point of view, however, it is necessary to scan over $a_0$ with a strong bias towards small values. This is because if one were to scan with say a uniform distribution on $a_0$, one would numerically have a hard time resolving the details of the EI ensemble for small $a_0$, which in the no-boundary state dominate probabilities for observations (Section \ref{probabilitysec}). Incidentally one can always revert to another distribution by weighting data points by an appropriate function of $a_0$.}. \\

According to the details described above, we probed the D-brane landscape by solving repeatedly the equations of motion \eqref{scalareqns} - \eqref{hdoteq} with the mentioned initial conditions and random model parameters. The outcome of a simulation can conveniently be categorised in three groups as was done in \cite{Agarwal2011,Dias2012}: either the brane succeeds in reaching the bottom ($x = 10 a_0$) of the throat - a succesful realisation of inflation - or it gets expelled from the throat ($x > 1$) in which case we are forced to discard the simulation because of our ignorance of the physics in the bulk, or it ends up in a metastable false vacuum (FV) minimum ($10 a_0 < x^* < 1$). We also discard the latter trials although it should be noted that they might alter some of our predictions. We plan to return to this point in future work (see also Section \ref{EI}). In total we performed $4 \cdot 10^9$ simulations, 99304 or $2.5 \cdot 10^{-3} \%$ of which were succesful realisations of inflation with more than 60 efolds of inflation\footnote{These numbers differ from \cite{Agarwal2011} because we sampled $a_0$ and $Q$ differently. Only background trajectories with a significant amount of inflation are relevant for TD probabilities (Section \ref{probobs}) however.}.

\section{Eternal Inflation near Inflection Points}
\label{EIsec}

\subsection{Inflation}
\label{inflation}

\begin{figure}[h!]
\centering
\includegraphics[width=5in]{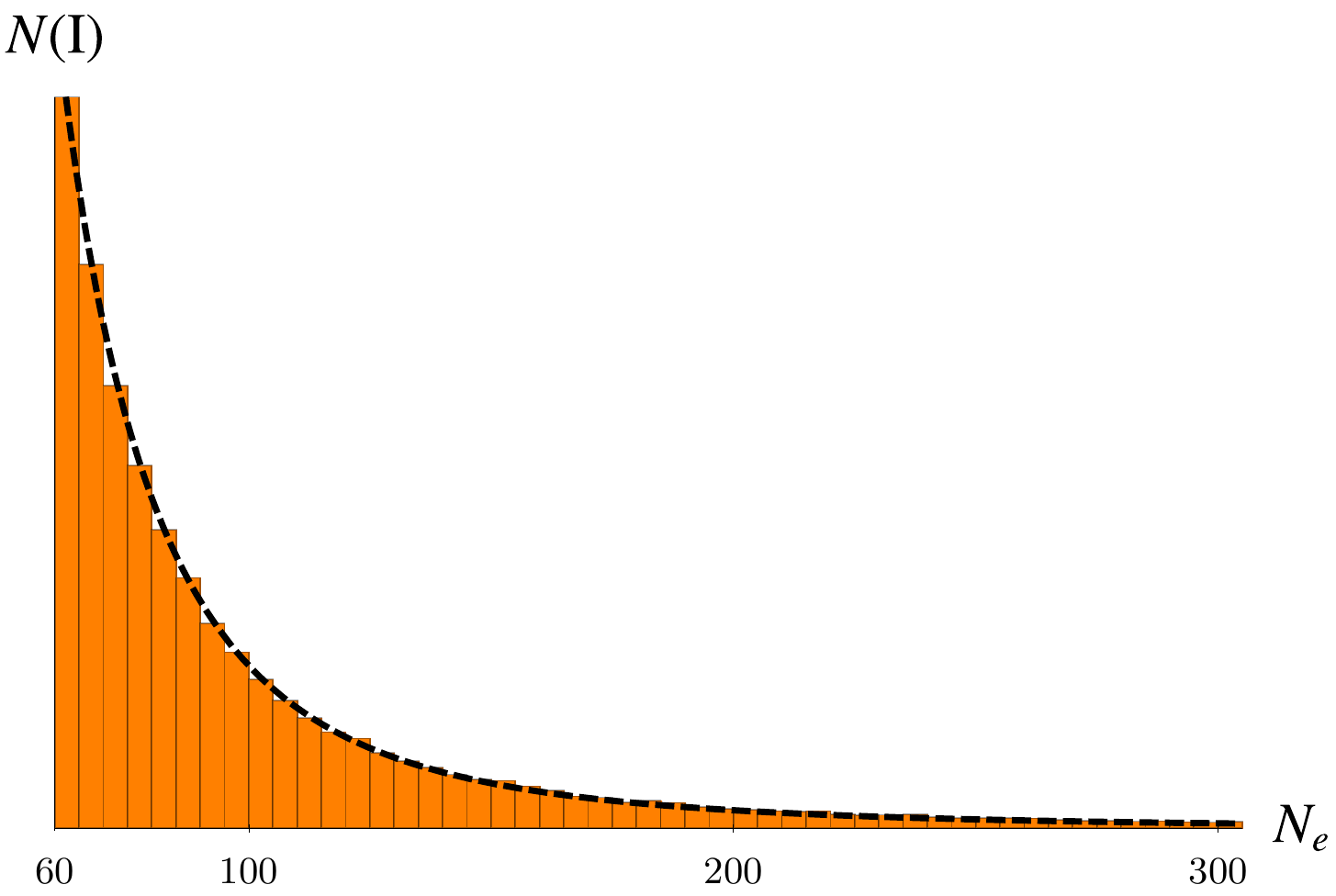}
\caption{Fraction of inflationary trajectories $N(\mbox{I},N_e)$ in our simulations with $N_e > 60$ efolds versus $N_e$. As in \cite{Agarwal2011}, we find the distribution is well-modeled by a power law $\propto N_e^{-\alpha}$ with $\alpha = 3.18 \pm .014$ at $2\sigma$. This roughly agrees with the distribution obtained in a model of inflection point inflation. The correspondence continues to hold for $N_e > 300$.}
\label{NIvsNe}
\end{figure}

\begin{figure}[h!]
\centering
\includegraphics[width=5in]{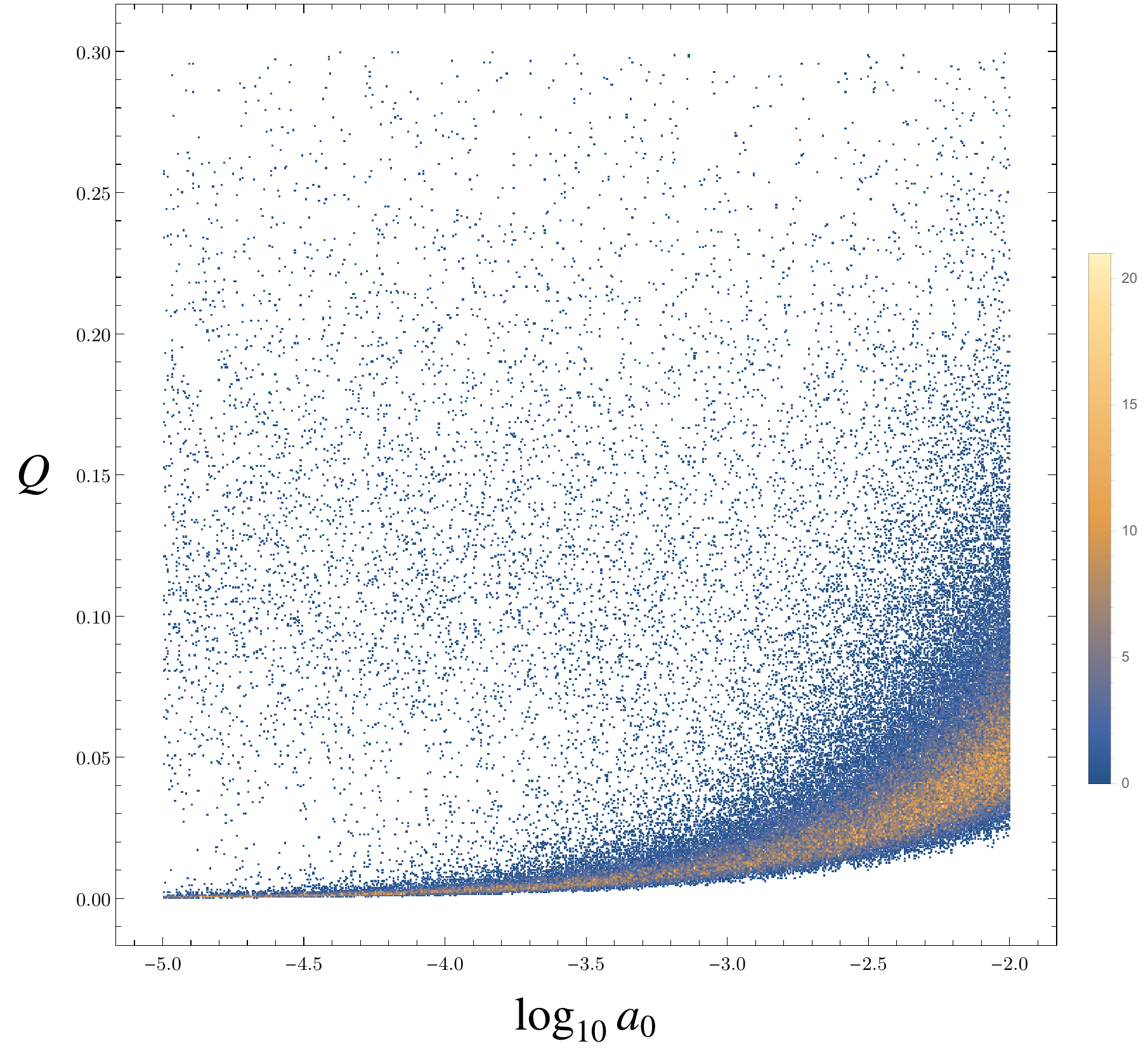} 
\caption{Distribution of trajectories with $N_e > 60$ in the $(\log_{10}a_0,Q)$-plane assuming a uniform measure when scanning $\log_{10}a_0 \in [-5,-2]$ and $Q \in [0,.3]$.}
\label{Qvsa0I}
\end{figure}

It is a well established fact that trajectories with an extended period of inflation typically begin by rapidly spiraling in the angular directions and moving downwards in the radial direction, until a sufficiently flat inflection point in the potential is encountered where slow roll occurs \cite{Agarwal2011,Dias2012}. After the slow roll phase, the brane falls off the Coulomb ledge and plummets towards the antibrane. Although radially directed motion is predominant during slow roll in the majority of cases, the motion along the inflection point can have significant angular components. In particular in general the mass-squared in the adiabatic (= along the motion) direction can become of the order of $H^2$ and the trajectory may exhibit a significant amount of bending \cite{Mcallister2012}. These features signal the generic invalidity of the single-field slow roll approximation to calculate properties of the scalar power spectrum, such as its tilt. In realisations of this kind one must resort to methods which explicitly track the evolution of perturbations around the homogeneous background in the multidimensional landscape. Slow roll-violating and multifield effects on CMB-related observables were extensively studied in \cite{Mcallister2012,Dias2012} to which we refer the reader for more details.

However, it was also found in \cite{Mcallister2012} that trajectories producing enough ($\gg$ 70) efolds of inflation do allow for an effectively single-field slow roll description. This is because the long phase of inflation before the observable scales in the CMB exit the horizon almost always\footnote{In 98.7\% of this ensemble.} suffices to damp out all non-slow roll and multifield effects. The phenomenology of the model is then that of inflection point inflation \cite{Baumann2007,Agarwal2011,Baumann2014}. We will make use of this result when we discuss perturbations in the ensemble of eternally inflating trajectories in Section \ref{probabilitysec}. Here, we recall that also the statistical distribution $N(\mbox{I},N_e)$ of inflating trajectories with $N_e> 60$ efolds as a function of $N_e$ in D-brane inflation behaves roughly as the distribution in inflection point inflation \cite{Agarwal2011}. Our result for the distribution, in accordance with \cite{Agarwal2011}, is illustrated in Fig. \ref{NIvsNe}.

In our work, we allow $\log_{10} a_0$ to vary over the interval $[-5,-2]$ and $Q$ to vary over the interval $[0;0.3]$, both uniformly, as detailed in Section \ref{ensembleconstr}. Fig. \ref{Qvsa0I} shows the statistical distribution of instances we found with $N_e > 60$ efolds of inflation in the $(\log_{10}a_0,Q)$-plane. A clear accretion of data points can be seen in a band in $Q$-space which becomes narrower and moves towards lower $Q$ as $a_0$ decreases. This can be understood from the picture of competing forces on the brane (Section \ref{branereview}): for fixed $Q$, as $a_0$ decreases the relative strength of the Coulomb force with respect to the flux forces diminishes. In the absence of an attractive term, the random forces typically either quickly eject the brane from the throat or cause it to reach the antibrane in just a few efolds. To restore balance to the force, $Q$ must also decrease. Obviously there are trajectories which do not obey this rule of thumb, namely those with large $Q$ and small $a_0$. They are anomalous in the sense that there is often no well-defined inflection point to be found along the path but still there is a long period of inflation. By coincidence the potential and initial conditions are just right to cause the brane to slowly spiral down to the tip, exploring a significant part of the angular manifold. To our knowledge such trajectories are a new finding, but because they turn out to be irrelevant for predictions (see Section \ref{EI}, in particular Fig. \ref{Qvsa0EI}) we will not study them in more detail here.

\subsection{Eternal Inflation}
\label{EI}

\begin{figure}[h!]
\centering
\includegraphics[width=5in]{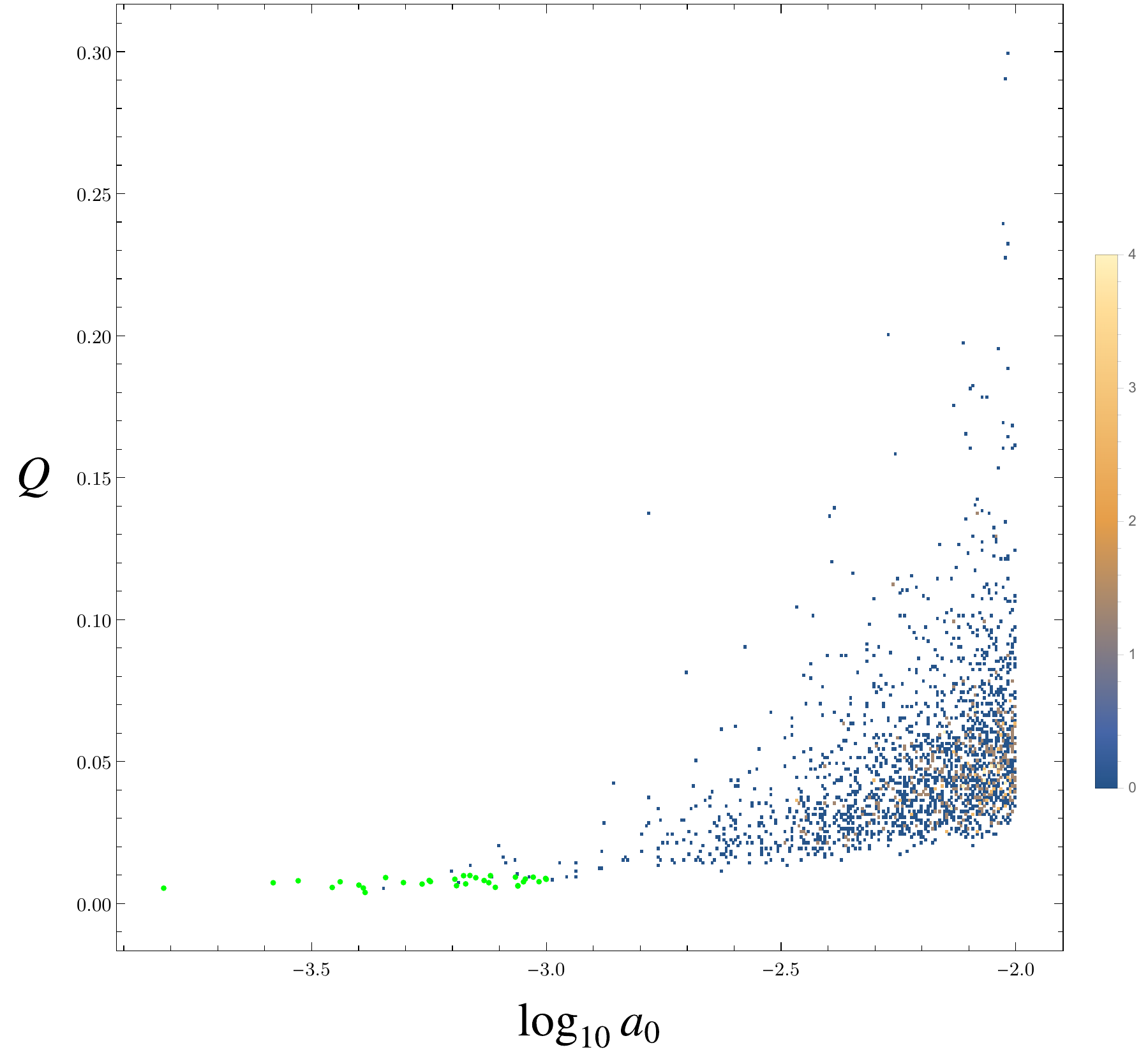} 
\caption{Distribution of EI trajectories in the $(\log_{10}a_0,Q)$-plane assuming a uniform measure on $\log_{10}a_0 \in [-5,-2]$ and $Q \in [0,.3]$. Green data points originate from a separate scan at small values of $a_0$ and $Q$ as discussed in the text.}
\label{Qvsa0EI}
\end{figure}

\begin{figure}[h!]
\centering
\includegraphics[width=5in]{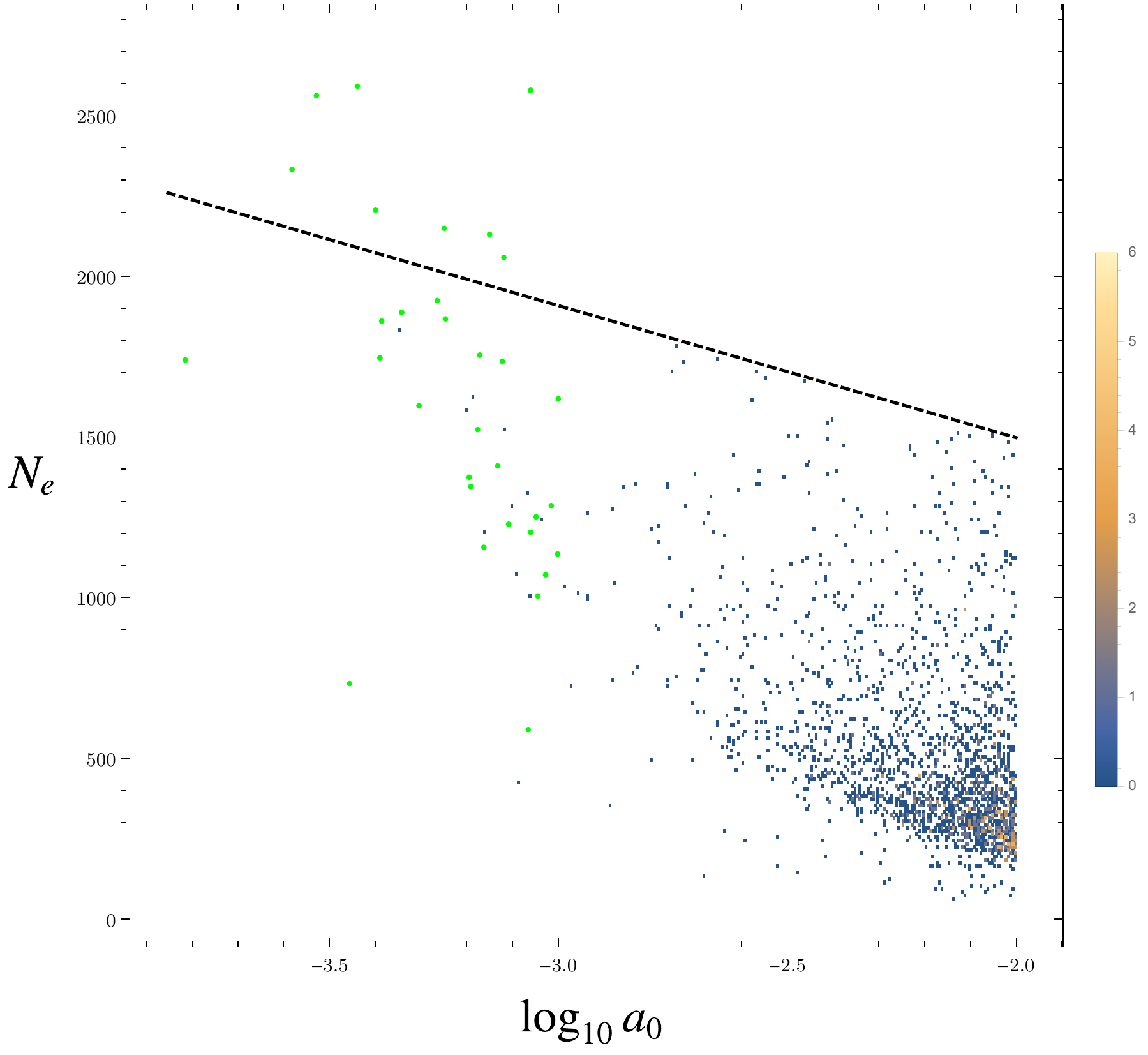} 
\caption{Distribution of $(\log_{10}a_0,N_e)$ for the EI ensemble. Here $N_e$ refers to the number of efolds of a \textit{background} trajectory that crosses an EI patch of the potential. The black line indicates the maximum number of efolds $N_{\tiny \mbox{max}}$ we are sensitive to in our simulations as a function of $a_0$ (see also Footnote \ref{Nmaxremark}). As $a_0$ decreases, the EI condition becomes increasingly difficult to satisfy: only the flattest fine-tuned potentials, which generate many efolds of inflation, make the cut. Green data points originate from a separate scan at small values of $a_0$ and $Q$, where $N_{\tiny \mbox{max}}$ was increased to 3000 (independent of $a_0$).}
\label{Nevsa0EI}
\end{figure}

\begin{figure}[h!]
\centering
\includegraphics[width=5in]{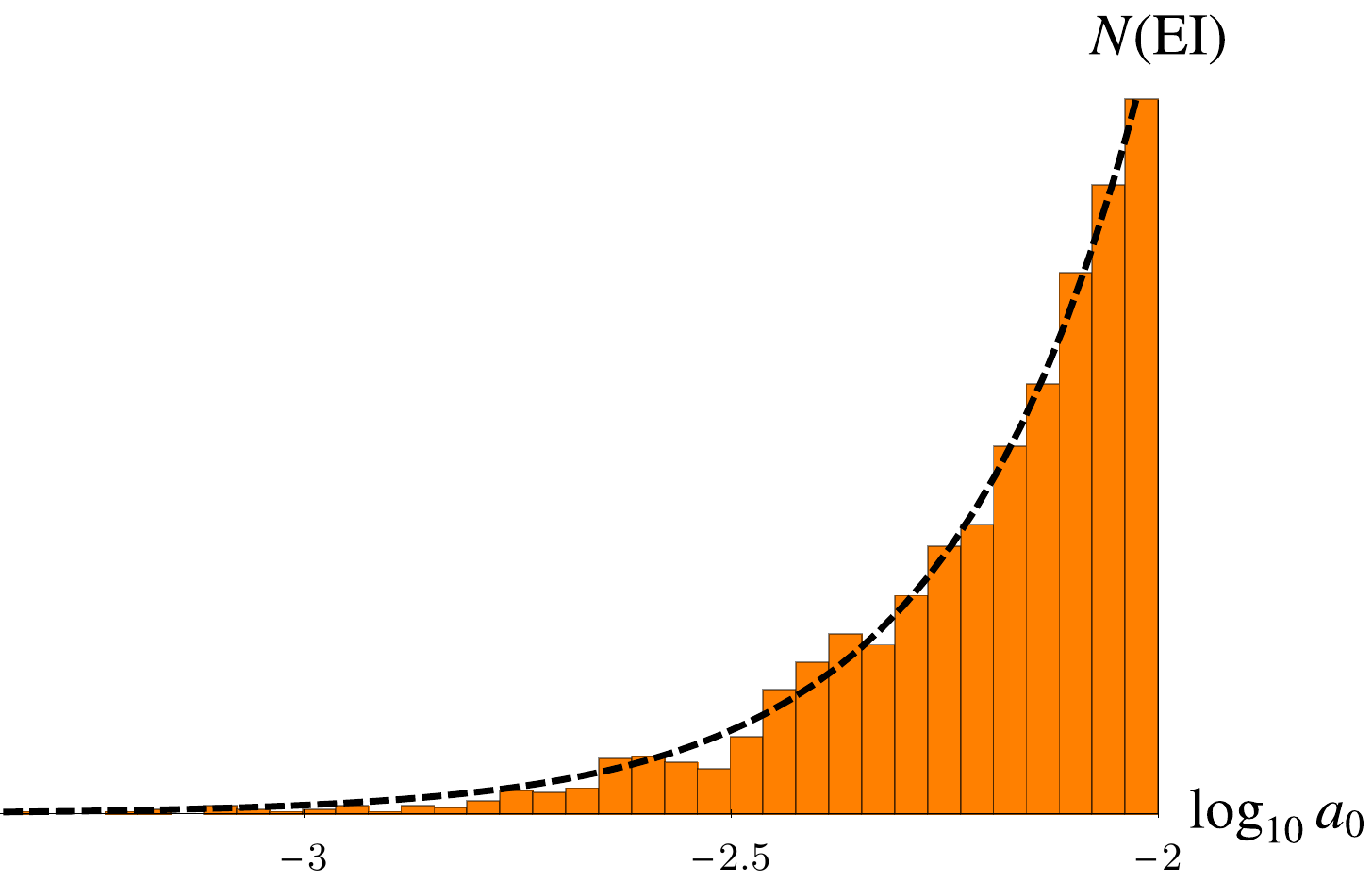} 
\caption{Fraction of EI trajectories in our simulations versus $\log_{10} a_0$. This is the information of Fig. \ref{Qvsa0EI}, marginalised over $Q$. We have fitted the data to an exponential of the form $\exp \left( \gamma \log_{10} a_0 \right)$ and found $\gamma = 4.53$ to be the best fit. This statistical information about the EI ensemble enters in the top-down probabilities for our observations, see Section \ref{amppert}.}
\label{NEIvsa0}
\end{figure}

\begin{figure}[h!]
\centering
\includegraphics[width=5in]{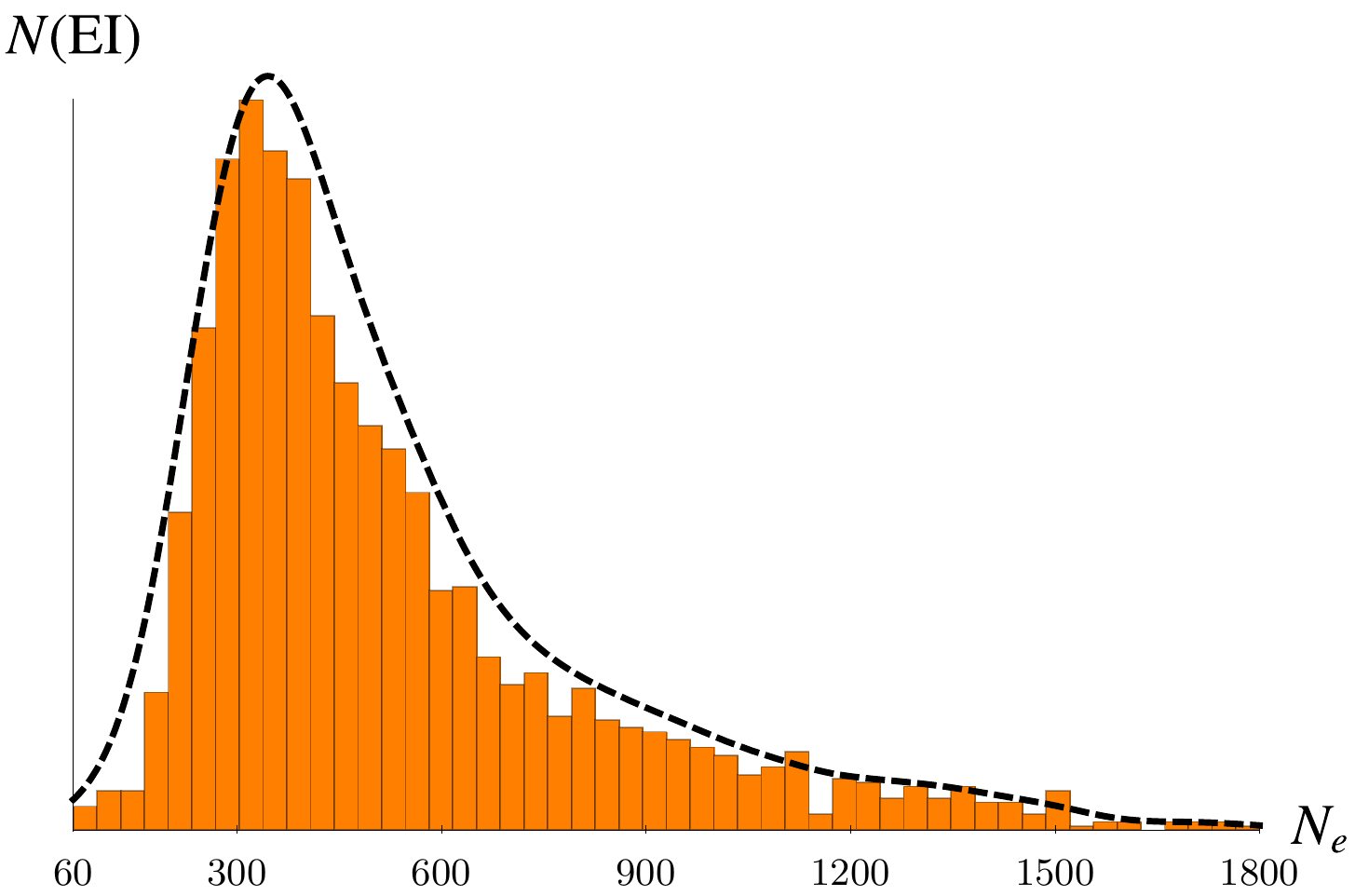}
\caption{Fraction of background EI trajectories $N(\mbox{EI},N_e)$ versus $N_e$. This is the information of Fig. \ref{Nevsa0EI}, marginalised over $Q$. Realisations with $\gtrsim 60$ efolds of classical inflation occur relatively less frequently in the EI ensemble than in the entire $\{ N_e > 60 \}$ ensemble (Fig. \ref{NIvsNe}) because the potentials patches are generally not flat enough to satisfy the EI condition. The position of the peak at $N_e \approx 300$ is sensitive to our choice of measure when scanning over $a_0$, as can be inferred from Fig. \ref{Nevsa0EI}, but the fact that the relative number of trajectories with $N_e \sim 60$ decreases compared to $N(\mbox{I},N_e)$ (Fig. \ref{NIvsNe}) is a generic feature.}
\label{NEIvsNe}
\end{figure}

Several approaches to quantifying the regime of EI in single-field models of inflation have been taken in the literature. A crude approach is explained in e.g. \cite{Guth2007}, which we briefly recall. During a Hubble time $H^{-1}$, an initial Hubble volume $H^{-3}$ filled homogeneously with a canonically normalised scalar field at value $\phi$ expands by a factor $e^3 \approx 20$. Quantum fluctuations $\Delta \phi_{\tiny \mbox{qu}}$ of the scalar field with respect to $\phi$, averaged over each of the twenty new Hubble volumes, may be approximated as independent\footnote{Correlations extend roughly over about a Hubble length due to the no-hair theorem for de Sitter space.} Gaussian distributed random variables with standard deviation $H/2\pi$ and mean zero. During the same time, the inflaton classically moves a distance $|\Delta \phi_{\tiny \mbox{cl}}| = |\dot{\phi}| / H$ down the potential hill. EI is then defined to occur when the probability $\Delta \phi_{\tiny \mbox{qu}} > |\Delta \phi_{\tiny \mbox{cl}}|$ is larger than $1/e^3$. This boils down to the condition
\be \label{EIGuth}
\dot{\phi}^2 < \frac{\left[\mbox{erfc}^{-1}(2/e^3)\right]^2}{2\pi^2} H^4 = \frac{C}{2\pi^2} H^4,
\ee
with $C \approx 1.36$. Other authors, e.g. \cite{Chen2006}, more simply model quantum fluctuations as a random walk with fixed step size $\Delta \phi_{\tiny \mbox{qu}} = H/2\pi$ per Hubble time, which leads to $C = 1/2$. A more careful analysis \cite{Creminelli2008} argues that one should take $C = 3$ instead of 1.36 or 1/2. These authors show that the probability to have an infinite reheating volume becomes nonzero (to first order in the slow roll parameters and $(H/\Delta \phi)^2$ where $\Delta \phi$ is the field range of the slow roll regime) if the condition \eqref{EIGuth} with $C=3$ is satisfied somewhere along the trajectory.

The above considerations apply to single-field inflation, whereas we know D-brane inflation is a multifield model where slow roll is generally violated somewhere along an inflating trajectory. During the bulk efold-time in extended cases of inflation the inflaton certainly is in slow roll, however, in the sense that $\epsilon, \sqrt{\epsilon}\eta^\parallel, \sqrt{\epsilon}\eta^\perp \ll 1$ which are the indicators of multifield slow roll inflation \cite{GrootNibbelink2000} (see Section \ref{probabilitysec} for a recollection of the definition of these quantities in the context of multifield inflation). Motivated by this, we propose the following minimal extension of the EI criterium \eqref{EIGuth} to determine whether a member of the ensemble of unperturbed, representative inflating trajectories in D-brane inflation has a regime of eternal inflation somewhere along the trajectory\footnote{Due to our choice of initial conditions $\dot{\phi}^A = 0$, \eqref{EImultifield} will always be satisfied near $t=0$. We obviously do not count this as a regime of EI.},
\be \label{EImultifield}
|\dot{\phi}|^2 < \frac{3}{2\pi^2} H^4.
\ee
Here $|\dot{\phi}|^2 \equiv G_{IJ} \dot{\phi}^I \dot{\phi}^J$ is the magnitude-squared of the velocity vector in inflaton field space, i.e. the singular conifold, and we have followed \cite{Creminelli2008} in our choice of $C$. This readily reduces, after a field redefinition, to \eqref{EIGuth} in the special case of effectively single-field inflation and incorporates that the inflaton field space is curved. Of course, \eqref{EImultifield} cannot be a valid criterium for eternal inflation in a general multidimensional potential landscape\footnote{See \cite{Dubovsky2011} for a rigorous approach.}. Indeed, if the classical trajectory would happen to be along a ridge on the potential hill, nearly any quantum fluctuation would destabilise it and inflation would certainly not be eternal. The actual perturbed paths would explore wholly different regions of the potential landscape than the classical background in this case, typically not be eternally inflating, and thus drastically alter predictions. However, instabilities of this kind do not occur in the  $\{ N_e > 60 \}$ ensemble of classically inflating trajectories described in Section \ref{inflation}. In fact, it turns out that when we condition on $> 60$ efolds of inflation, the masses of the entropic fields are typically much greater than the adiabatic mass which becomes tachyonic when enough ($\gg 100$) efolds of inflation are realised \cite{Mcallister2012, Dias2012}. We have reverified this claim for our ensemble of inflating trajectories by calculating the eigenvalues $m_i^2, i = 1...6$ of the mass matrix
\be
\tensor{M}{^I_J} = \tensor{V}{^{;I}_J} - \tensor{R}{^I_{KLJ}}\dot{\phi}^K \dot{\phi}^L - \frac{1}{a^3} \mathcal{D}_t \left[ \frac{a^3}{H} \dot{\phi}^I \dot{\phi}_J \right],
\ee
where $V_{;IJ} \equiv \partial_{I} \partial_J V - \Gamma^K_{IJ}\partial_K V$ and find agreement with \cite{Mcallister2012, Dias2012} to which the reader is redirected for more details. Because the potential is strongly convex in directions orthogonal to the classical path during slow roll, we do not expect quantum fluctuations to destabilise it. Notice that the multifield EI condition \eqref{EImultifield} can be rewritten by using the slow roll approximation $|\mathcal{D}_t \dot{\phi}^A| \ll |3H\dot{\phi}^A|, |G^{AB} \partial_B V|$ and $|\dot{\phi}|^2 \ll V$ in \eqref{scalareqns} - \eqref{hdoteq},  from which one deduces $|\dot{\phi}|^2 \approx |\partial V|^2 / 9H^2$ and $H^4 \approx V^2/9$. One obtains
\be \label{EIapprox}
2 \pi^2 \frac{|\partial V|^2}{V^2} < V.
\ee
This restores the intuition that sufficiently flat patches of the potential are eternal inflation patches. In terms of the slow roll function $\epsilon \equiv -\dot{H}/H^2$, we have, in the slow roll approximation, $\epsilon \approx |\partial V|^2 / 2V^2$. In this way we recover the EI condition ``$V > \epsilon$" that was mentioned in the introduction. \\

We now turn to the statistical distribution of eternally inflating trajectories in D-brane inflation, which we identify with eternal inflation patches of the potential landscape. The condition $V > \epsilon$ implies that the scalar amplitude is of order 1 on scales associated with eternal inflation. Hence one expects that the most interesting trajectories, i.e. those with a small scalar perturbation amplitude on observable scales, will typically have a very long period of non-eternal slow roll inflation after the exit from eternal inflation. This turns out to be always the case; trajectories with EI have $N_e \gg 60$ of slow roll, which in turn implies they are effectively single-field\footnote{We also checked explicitly that our EI trajectories allow for an effectively single-field description by using the transport method and verifying that the difference with the slow roll result is negligible. For this we used a code which is publicly available at \href{http://transportmethod.com}{\texttt{transportmethod.com}} and described in \cite{Dias2015}.}. This means our EI condition \eqref{EImultifield} is consistent in the landscape of D-brane inflation, but we do not have a proof that it is generally sufficient\footnote{This would require checking the condition (47) of \cite{Dubovsky2011}, a herculean task for this model.}.

Fig. \ref{Qvsa0EI} shows the distribution of EI trajectories in the $(\log_{10}a_0,Q)$-plane: out of the 99304 inflationary realizations with $N_e > 60$, only 2106 or $\sim 2.1\%$ possess an EI patch. It is clear that EI occurs only extremely rarely for small values of $a_0$. Actually below $\log_{10} a_0 \sim -3.8$ our $\sim 10^9$ trials did not yield any EI realisations even though we have every reason to expect EI patches continue to exist for smaller values of $a_0$ yet.

This can be understood from the rough $V > \epsilon$ criterium, which simply becomes harder to satisfy when $a_0$ decreases because the potential is required to be so flat\footnote{The effect that our choice of measure in $Q$-space also tends to lower the number of $N_e > 60$ trajectories found (cf. Fig. \ref{Qvsa0I}) is negligible in comparison.} (recall that we take $V \sim a_0^4$). We believe this difficulty may slightly be alleviated by using a more complete form of the potential that retains terms with higher powers of $x$ \cite{Agarwal2011,Dias2012,Mcallister2012}. This is because the number of Wilson coefficients rises greatly with more terms, plausibly allowing for more fine-tuning in the potential landscape. However, when doing this, each simulation becomes computationally more expensive. We made a trade-off between time cost and accuracy.

Because the rare EI realisations with small $a_0$ contribute significantly to the probabilities for observations (cf. Section \ref{NBWF}), we performed a separate scan of the potential landscape in a very restricted region of parameter space at relatively small values of $a_0$ and $Q$ in order to resolve this corner as best as we could given our computational limitations. We mark the data points we found in this corner by green circles in the scatter plots Figs. \ref{Qvsa0EI}, \ref{Nevsa0EI}, \ref{Asvsa0EI}, \ref{nsvsa0EI}, \ref{alphasvsa0EI} and \ref{rvsa0EI} to emphasize they were obtained by a different scanning measure on $a_0$ and $Q$ as the other data points (Section \ref{ensembleconstr}). The extra data points are not included in the histograms Figs. \ref{NEIvsa0} and \ref{NEIvsNe}.

Next, it is illuminating to look at the distribution of EI data points in the $(\log_{10}a_0,N_e)$-plane. This is shown in Fig. \ref{Nevsa0EI}. A rough explanation of the general behavior can again be given: because the potential is required to be flatter and flatter to satisfy the EI condition as $a_0$ decreases, histories inflating along these flat patches are expected to produce many efolds of inflation. We emphasise that this is only a rule of thumb\footnote{\label{Nmaxremark} A technical remark: when simulating the dynamics one must specify a maximum efold-time $N_{\tiny \mbox{max}}$, after which one assumes that the brane has ended up in a FV if it has not by then been ejected from the throat or reached the threshold value near the tip. Because the time cost of a simulation rises rapidly with $N_{\tiny \mbox{max}}$, and guided by the rule of thumb, we made $N_{\tiny \mbox{max}}$ a function of $a_0$ to most effectively detect EI. We will have missed some instances with many efolds, but again a trade-off between time cost and accuracy is unavoidable.}. Two more pieces of statistical information, which visualise part of the content of Figs. \ref{Qvsa0EI} and \ref{Nevsa0EI} in another way, are given in Figs. \ref{NEIvsa0} and \ref{NEIvsNe} respectively.

One might wonder whether we have identified a representative set of patches of eternal inflation in the D-brane landscape so that our predictions are robust. Actually, because of the flux-induced terms, there are also patches of false vacuum eternal inflation in the throat. This leads to the question whether there are EI histories which undergo a prolonged period of slow roll inflation after tunnelling out of the false vacuum. If histories of this kind exist they may well compete in the NBWF with the eternally inflating trajectories of the slow roll type we considered, and therefore alter our predictions e.g. for the universe being open or closed.
In the $\sim 10^{6}$ inflationary landscape patches we have generated so far it appears that even though false vacua are relatively common, tunnelling in the radial direction is not at all. In particular we have never found a slow roll patch across the Coulomb barrier. Instead after tunnelling inflation always ends within a fraction of an efold. However to show conclusively that slow roll EI patches dominate TD probabilities we should verify that tunnelling does not occur from a false vacuum in a general $(r,\Psi)$-direction, to a region where the slow roll conditions hold. One should also examine more complete forms of the potential which retain higher powers of $x$, which may possess the necessary additional structure for false vacuum tunnelling plus slow roll. We leave a more detailed analysis of this for future work.

\section{The Quantum State} 
\label{NBWF}

The statistical ensemble of inflating backgrounds in the D-brane landscape exhibits a broad range of values for cosmological observables. This leads to broadly distributed probabilities and therefore weak predictions if one adopts a flat prior on initial conditions in the landscape, reflecting ignorance about the underlying theory of the quantum state of the universe. We take a different viewpoint and instead complete our theoretical framework with a definite model of initial conditions, the NBWF, which we discuss here. In Section \ref{probabilitysec} we show the NBWF applied to the D-brane landscape leads to sharply peaked probabilities for cosmological observables.

\subsection{No-Boundary Wave Function}
\label{NBWFsec}

A quantum state of the universe is specified by a wave function $\Psi$ on the superspace of three-geometries ($h_{ij}(x)$) and matter configurations ($\phi(x)$) on a closed, spacelike three-surface $\Sigma$. Schematically we write $\Psi=\Psi[h_{ij},\phi]$ where $\phi$ stands for a higher-dimensional scalar field. A wave function of the universe $\Psi[h_{ij},\phi]$ does not select a single universe. Rather it describes an ensemble of possible histories of the universe, in different patches of the landscape potential, along with their quantum probabilities.

We adopt the no-boundary wave function (NBWF) \cite{Hartle1983} as a model of this state. In the semiclassical approximation the NBWF is given by
\be
\Psi[h_{ij},\phi] \approx  \exp(-I[h_{ij},\phi]) = \exp(-I_R[h_{ij},\phi] +i S[h_{ij},\phi])
\label{semiclass}
\ee
where $I_R[h_{ij},\phi]$ and $-S[h_{ij},\phi]$ are the real and imaginary parts of the Euclidean action $I$ of a compact {\it regular} saddle point solution that matches the real boundary data $(h_{ij},\phi)$ on its only boundary $\Sigma$. If $S$ varies rapidly compared to $I_R$ the wave function takes a WKB form and predicts that the boundary configuration $(h_{ij},\phi)$ evolves as a Lorentzian, classical universe. The ensemble of classical universes predicted by the NBWF is of particular interest in cosmology, since classical spacetime is a prerequisite for cosmology as we know it. Indeed our existence {\it selects} the quasi classical realm of the wave function of the universe. The classical ensemble predicted by the NBWF has the striking property that it is peaked around cosmological backgrounds which undergo some amount of matter driven, slow roll inflation \cite{Hartle2007,Hartle2008}. Hence the NBWF populates the landscape in a very specific manner: it {\it selects} the landscape patches where the conditions for inflation hold and within those patches, it selects universes with an early period of inflation.

The tree-level probabilities $P_{NB}(\phi_0)$ of the individual background histories in the NBWF ensemble are proportional to $\exp[-2 I_R(h_{ij},\phi)]$ and therefore specified by the real Euclidean part of the saddle point. It turns out that to an excellent approximation these are proportional to \cite{Hartle2008}
\be
P_{NB} \propto \exp[-2 I_R] \approx \exp \left[\frac{3\pi}{\Lambda + V(\phi_0)}\right]
\label{bu}
\ee
where $\phi_0$ is the value of the scalar field at the start of inflation and we have included a small cosmological constant, which we will assume may vary over the set of stable flux compactifications such that it is (statistically) uniformly distributed over the narrow anthropically allowed range. The probabilities \eqref{bu} are conserved during classical evolution, and in particular during inflation, as a consequence of the Wheeler-DeWitt equation.

The probabilities for scalar curvature perturbations $\zeta$ and tensor perturbations $t_{ij}$ on $\Sigma$ are specified by the no-boundary wave function of perturbations around homogeneous saddle points \cite{Halliwell1985,Hartle2010b}. In the semiclassical approximation these are given by the usual product\footnote{The regularity condition on the saddle points implies that perturbations start out approximately in the Bunch-Davies ground state \cite{Halliwell1985}.} of nearly Gaussian probabilities $P(\zeta_k|\phi_0)$ and $P(t_{ij(k)}|\phi_0)$ for fluctuation modes $\zeta_k$ and $t_{ij(k)}$ on $S^3$ \cite{Halliwell1985,Hartle2010b}. The statistical properties of these distributions - which are the usual observables in inflationary cosmology -  are given in terms of properties of the backgrounds. We return to these below in Section \ref{probabilitysec} in the D3-brane landscape.

We now apply the no-boundary weighting to the D-brane landscape. Since the NBWF selects inflationary universes it suffices to limit attention to inflationary patches of the landscape. This is precisely what we have done in the previous Sections, where we have retained the choices of Wilson coefficients and warp factor at the tip, $\{c_{LM}, a_0\}$, that give rise to inflationary trajectories. The trajectories that we identified are effectively single-field\footnote{At least when $N_e \gg 70$. We will see that other trajectories hardly contribute to the NBWF probabilities for observables.} and start high in the throat, thus probing the entire inflationary patch specified by $\{c_{LM},a_0\}$. However the NBWF also allows for inflationary histories that start at a lower value in the same patch. The ensemble of inflationary universes predicted by the NBWF in the D-brane landscape can therefore be labeled by $\phi_0^J$ where $J \equiv \{c_{LM},a_0\}$ specifies the patch and $\phi_0$ is the initial value of the scalar field in the effectively one-dimensional potential. It is useful to distinguish between two scales in the problem. The first is the overall scale of inflation, set by $a_0$ through $D_0 = 2 T_3 a_0^4$. The second is the difference $\Delta V / V \ll T_3 = 10^{-2}$ between the value of the potential at both ends of a particular inflationary patch. Thus when comparing the no-boundary weight \eqref{bu} for histories associated with widely different scales $a_0$, more specifically when $\Delta \log_{10} a_0 \gtrsim 10^{-3}$, we may approximate
\be \label{buapprox}
P_{NB}(\phi_0^J) \approx  \exp \left[\frac{3\pi}{2 T_3 a_0^4}\right],
\ee
while we must retain the full form \eqref{bu} when comparing the relative probabilities assigned by the NBWF to histories starting at different values $\phi_0$ in the same patch $J$. When perturbations are included, saddle points with $V < \epsilon$ at $\phi_0^J$ give rise to an ensemble of nearly homogeneous inflationary universes with small fluctuations. By contrast, saddle points starting in eternal inflation regions of the landscape where $V > \epsilon$ predict high amplitudes for significantly inhomogeneous universes that have large (very) long-wavelength perturbations on the scales associated with eternal inflation \cite{Hartle2010b}. This is because in eternal inflation, the typical size of perturbations on the horizon scale is comparable or larger than the background field motion in a Hubble time. As a consequence the perturbations have a large effect on the structure of the universe on those scales. This is important when it comes to probabilities for our observations as we discuss next.

\subsection{Probabilities for Observations}
\label{probobs}

So far we have discussed the NBWF probabilities for members of the ensemble of histories of the whole universe. Probabilities relevant for the prediction of our observations however are conditional probabilities that describe correlations between cosmological observables ${\cal O}$ given our observational situation. We have called the former bottom-up (BU) probabilities and the latter top-down (TD) probabilities \cite{Hawking2006}. Here we briefly review how one gets from BU probabilities for histories to TD probabilities for observations. For details we refer the reader to \cite{Hartle2013,Hertog2013}. 

We model our observational situation in terms of a set of ultra local data $\Dobs$ on the scales of our planet. All we know is that our universe exhibits at least one instance of $\Dobs$ ($\Dobsaoi$ for short) --- our instance. The TD probability for us to observe $\cO$ is given by the BU probabilities for histories with $\cO$ conditioned on the requirement that each history contains at least one instance of $\Dobs$ (in a region where the observable takes the value $\cO$),
\be
P_{TD}(\cO) = P(\cO|\Dobs^{\ge 1}) .
\label{TD}
\ee
In the D-brane landscape this takes the form
\be
\label{Pobs}
P({\cal O}| \Dobsaoi) \propto \sum_{J = \{c_{LM},a_0\}} \int \mathrm{d}\phi_0^J \hspace{2mm} P(\Dobsaoi|{\cal O},\phi_0^J) P_{NB} ({\cal O}|\phi_0^J) P_{NB}(\phi_0^J).
\ee

The last two factors are the NBWF probabilities for observables and backgrounds. We will be interested in observables associated with statistical properties of the pattern of CMB perturbations. Those can be expressed in terms of the shape of the potential evaluated around the field value probed by the background when the associated scale crosses the horizon during inflation. This means the probabilities for $\cO$ do not explicitly depend on the perturbation variables $\zeta_k^J$ and $t_{ij(k)}^J$. We have therefore coarse-grained (summed) over the probabilities for the individual perturbed histories\footnote{Thus the NBWF probabilities $P_{NB}(\phi_0^J)$ in \eqref{Pobs} should be viewed as the {\it sum} of the probabilities of an ensemble of perturbed, inflationary histories which differ in their actual CMB patterns but share the same statistical features.} in \eqref{Pobs}.

The first factor in \eqref{Pobs} is the `top-down' weighting. The TD requirement that our observational situation exists somewhere in the universe is a global condition that selects from among the NBWF ensemble of possible histories ones in which $\Dobs$ occurs at least once, somewhere, sometime. This is of course a trivial condition in sufficiently large universes. After all, in an infinite universe anything occurs somewhere sometime with probability one. In universes that are small however the TD factor in \eqref{Pobs} is exceedingly small. This is because as observers we are physical systems within the universe, described by local data $\Dobs$, which occur only with a very small probability $p_E$ in any Hubble volume on a surface $\Sigma$ of approximately constant density. In universes with $p_E(\Dobs) \ll 1/N_h$, where $N_h(\phi_0^J)$ is the total number of Hubble volumes in $\Sigma$, the TD factor in \eqref{Pobs} is approximately $p_E N_h \ll 1$ and thus amounts to weighting the BU probabilities by the volume of $\Sigma$ \cite{Page1997,Hartle2009}. We have argued \cite{Hartle2009} this is the case for histories without a regime of eternal inflation, which have high amplitudes for nearly homogeneous universes only, with Vol$(\Sigma) \propto e^{3N_e}$. By contrast, histories with a regime of eternal inflation have high amplitudes for configurations that have large perturbations on scales associated with eternal inflation \cite{Hartle2010}. As a consequence the volume of $\Sigma$ of such universes is generally exceedingly large or even infinite so that the TD factor in \eqref{Pobs} is approximately equal to one. 

Hence the TD factor in \eqref{Pobs} strongly {\it favours} landscape patches $\{c_{LM},a_0\}$ where the conditions for eternal inflation hold. In those patches the low no-boundary probability of histories starting above the threshold for eternal inflation is more than compensated for by the large TD weighting. The net effect of the TD factor is therefore that the sum over $\{c_{LM},a_0\}$ in \eqref{Pobs} can be restricted to eternal inflation patches\footnote{Eternally inflating histories are highly inhomogeneous on the largest scales. However we will assume that their properties on observable scales are statistically the same in every Hubble volume in each individual history. That is, we coarse grain over quantum tunnelling events between different landscape patches. We hope to include the effect of bubble nucleations in future work.},
\be
\label{Pobs1}
P({\cal O}| \Dobsaoi) \propto \sum_{J_{EI} = {\{c_{LM},a_0\}_{EI}}} \int \mathrm{d}\phi_0^{J_{EI}} \hspace{2mm} P_{NB} ({\cal O}|\phi_0^{J_{EI}}) P_{NB}(\phi_0^{J_{EI}})
\ee
where ${\{c_{LM},a_0\}_{EI}}$ labels the eternal inflation patches, and the integral is restricted to starting points $\phi_0^{J_{EI}} > \bar \phi_0^{J_{EI}}$ where $\bar \phi_0^{J_{EI}}$ is the threshold of EI closest to the tip. This is independent of $p_E$ and of the precise nature of $\Dobsaoi$ and hence computable \cite{Hartle2009}.

Finally, within a fixed patch $J_{EI}$ the integrand in \eqref{Pobs1} is strongly peaked at $\phi_0^{J_{EI}} = \bar \phi_0^{J_{EI}}$ due to the heavily biased $P_{NB}(\bar \phi_0^{J_{EI}})$ factor. Hence
\be
P({\cal O}| \Dobsaoi) \propto \sum_{J_{EI}}  P_{NB} ({\cal O}|\bar \phi_0^{J_{EI}}) P_{NB}(\bar \phi_0^{J_{EI}})
\label{res}
\ee
where, because we sum over a large range of values of $a_0$, the latter factor may be approximated by \eqref{buapprox}.
Thus the NBWF favours low values of $a_0$ or, equivalently, low-lying eternal inflation patches. 

To summarise, the NBWF implies a prior for the prediction of local observations that selects landscape regions of eternal inflation and specifies a relative weighting of such regions that favours patches where eternal inflation occurs at a low value of the potential.

\subsection{Local Observables}
\label{testing}

Theories are tested, supported, and falsified by comparing their predictions with observation. What is important for this are the top-down probabilities for local observables conditioned on a description of the observational situation and coarse-grained over what is not observed.

An objective program for supporting a theory $\FT$ in cosmology consists of searching among all possible TD correlations in our total data $D$ on all scales (including the possible results of future observations) for ones that are predicted by $(I,\Psi)$ with significant probability and are sensitive to the theory's details \cite{Hartle2007a,Hartle2013}. The observables $\cO$ participating in the correlation must be selected to this end. Not every correlation will test the theory but we posit no list of correlations that must have significant probability for a theory to be successful\footnote{Evidently one theory will be regarded as more successful than another if it predicts more correlations.}. Instead we search through all.

We now investigate the predictions of the NBWF in the D-brane landscape for correlations among a range of observables related to the pattern of CMB perturbations, the cosmological constant $\Lambda$ and other data $\Dell$ describing the nearby universe in our Hubble volume, conditioned on the requirement that the universe contains our ultralocal observational situation $\Dobs$. That is, we calculate $P(\cO| \Dobsaoi)$ with $\cO = (\Lambda,A_s,n_s,\alpha_s,r, \Dell)$ where $\Dell$ are data that include our local distribution of galaxies which, together with $\Lambda$, specify a FLRW model and determine our time $t_0(\Dell,\Lambda)$ as measured, say, from reheating. Similar joint TD probabilities were considered previously in \cite{Hartle2013,Tegmark2006}.

The TD probabilities for $\cO$ are approximately given by \eqref{res}. From the definition of conditional probabilities it follows that
\be
\label{bayes}
P({\cal O}| \Dobsaoi) \propto \sum_{J_{EI}} P (\Dell | \Lambda,A_s, n_s, \alpha_s, r, \bar \phi_0^{J_{EI}}) 
P_{NB} (\Lambda,A_s, n_s,\alpha_s, r, \bar \phi_0^{J_{EI}}).
\ee
The first factor acts as a further selection on the ensemble because it will be proportional to the number of habitable galaxies that have formed in our Hubble volume by the time $t_0$. We approximate this by the expected number of galaxies for given values $(\Lambda,A_s)$, which follows from the classical equations of motion, and we include our data on the age $t_0$ in a similar manner by requiring that $t_0$ is greater than the classical collapse time $t_C(\Lambda,A_s)$ of a typical fluctuation. This gives
\be
\label{fractgal}
P (\Dell | \Lambda,A_s, n_s,r,\alpha_s, \bar \phi_0^{J_{EI}})  = P_g(\Dell) f(\Lambda, A_s, t_0) \theta(t_0-t_C(\Lambda,A_s)).
\ee
Here, $\theta(x)$ is the step function,  $f(\Lambda, A_s, t_0)$ is the fraction of protons in our Hubble volume in the form of galaxies by the time $t_0$, and $P_g(\Dell)$ is the probability that any one galaxy has the specific features of ours contained in the data $\Dell$. This is plausibly independent of $\Lambda$ and $A_s$ and, like $p_E$ above, will not appear in the final probabilities.

\begin{figure}[t]
\includegraphics[width=5in]{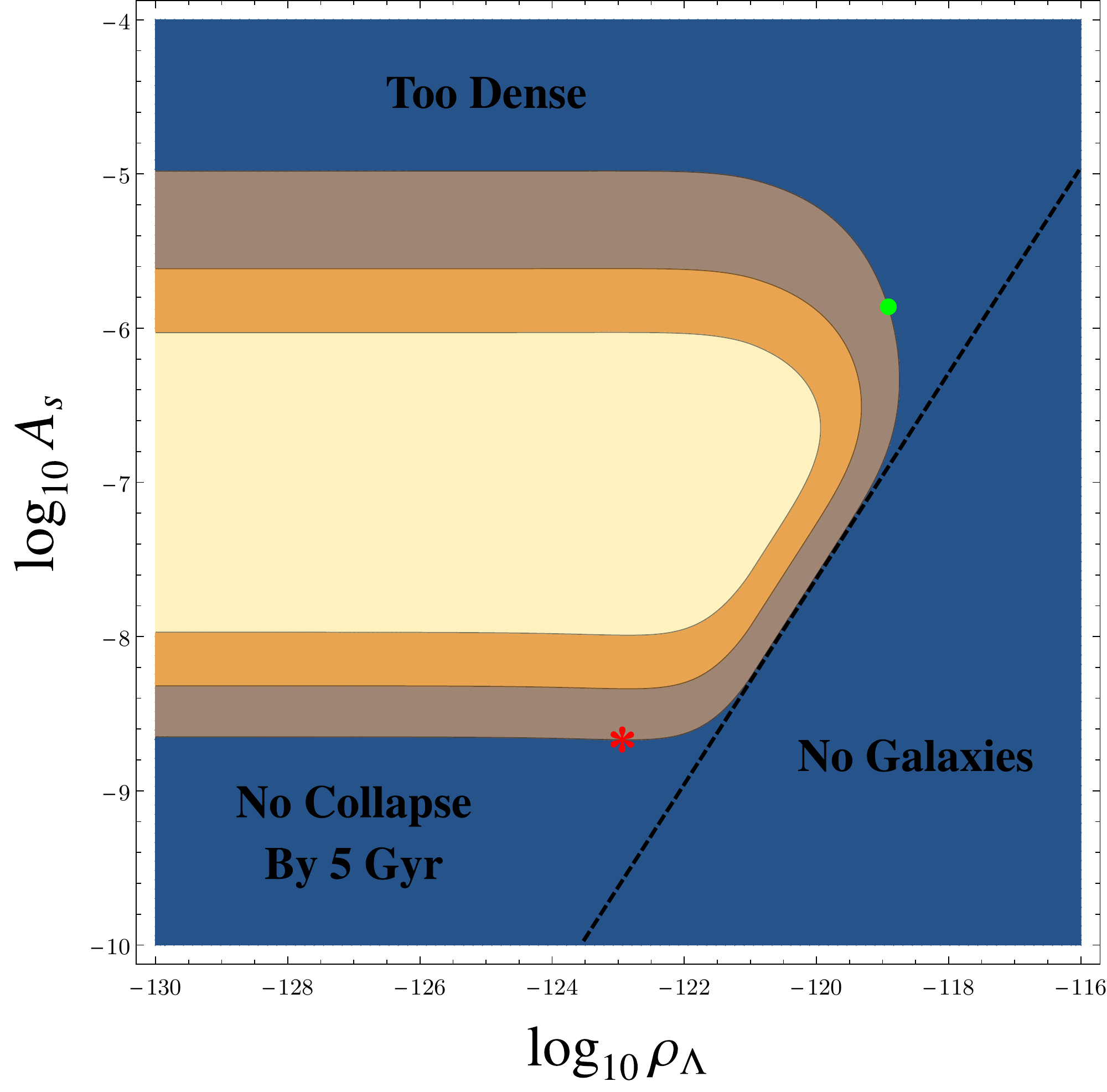}
\caption{The ($\Lambda$-$A_s$)-plane. This figure, adapted from \cite{Hartle2013,Tegmark2006} shows the various factors affecting the calculation of the probability $P(\Dell|\Lambda, A_s)$ that selects habitable histories. The region that allows for gravitationally collapsed bodies at some time is to the left of the diagonal dotted line assuming the presently observed matter density. The contours trace constant values of the fraction $f(\Lambda, A_s, t_0)$ of protons in galaxies at a time $t_0 \sim 5$ Gyr, a few billion years before the present age, as calculated from formulae in \cite{Tegmark2006}. The curves shown correspond to $f=.18,.36,.54$. The probability $P(\Dell|\Lambda, A_s)$ is approximately uniform in the interior of the contours shown and decreases quickly outside them. The most probable values for $\Lambda$ and $A_s$ are indicated by the {\color{red}\bigstar} assuming the NBWF. These coincide with the observed values. The correlation thus supports the theory that led to it.  The most probable values for a uniform prior in $A_s$ and $\Lambda$ are at the largest value of $\log_{10} \Lambda$ and $\log_{10} A_s$ in the allowed range and indicated by the {\color{green}\bigdot} and are not close to the observed values.}
\label{fraction}
 \end{figure}

Fig. \ref{fraction} shows a contour plot of $f(\Lambda, A_s, t_0)$, calculated along the lines of \cite{Hartle2013,Tegmark2006}, for $t_0 \sim 5$ Gyr, a few billion years before the present age to allow time for biological evolution. The fraction involves the combination $\Lambda/A_s^{3/2}$ (we fix the matter density per CMB photon $\xi$ to its observed value). This gives rise to the dotted line with slope 2/3 in Fig. \ref{fraction}. For combinations $(\Lambda, A_s)$ to the right of this line the universe expands too quickly for galaxy-sized, gravitationally bound systems to ever form. For $A_s \gtrsim 10^{-5}$ halos consist mostly of massive black holes, and the fraction $f$ of matter in habitable galaxies is essentially negligible.  
The fraction $f$ is also a function of time. For times much before the collapse time $t_C(\Lambda, A_s)$ the fraction $f(\Lambda, A_s, t) $ of protons in galaxies is very small. For times that are greater than $t_C(\Lambda, A_s)$ the fraction approaches an asymptotic value. The sharp constraint on the age $t_0$ represented by the step function in \eqref{fractgal} 
can thus be approximately implemented by a sharp constraint that the fraction $f$ be above some value $f_C$ by $t_0$. This in turn leads to a sharp lower cutoff on $A_s \geq 10^{-8.67} \equiv A_s^c$, and a sharp prediction of $\Lambda$ in agreement with observation \cite{Hartle2013,Planck_inflation}.

In the next Section we substitute the NBWF prior \eqref{bu} in the TD probabilities \eqref{bayes} to derive our predictions for the CMB-related observables in $\cO$ in D-brane inflation.

\section{NBWF Predictions for CMB Observables}
\label{probabilitysec}

In Section \ref{EI} we argued that the eternal inflation patches allow for an effectively single-field slow roll description on observable scales that leave the horizon well after the exit from eternal inflation. In this approximation the observables $A_s, n_s, \alpha_s$ and $r$ of interest are given by (e.g. \cite{GrootNibbelink2000})
\begin{align}
A_s &= \frac{V}{24 \pi^2 \epsilon}, \\
n_s &= 1 - 2 \eta^\parallel - 4\epsilon, \label{nsformula} \\
\alpha_s &= - \frac{\mathrm{d} n_s}{\mathrm{d} N_*}, \\
r &= 16 \epsilon,
\end{align}
which must be evaluated at around $N_* \approx 60$ efolds before the end of inflation. Here $\epsilon = -\dot{H}/H^2 \approx |\partial V|^2 / 2V^2$. In multifield inflation, $\eta$ is a vector with a component $\eta^\parallel$ along the trajectory and a component $\eta^\perp$ perpendicular to it, which quantify respectively the degree of acceleration of the brane parallel to its instantaneous motion and the degree of instantaneous bending of the trajectory on itself\footnote{In the EI ensemble, $95\%$ of trajectories satisfy the rough single-field condition proposed in \cite{Agarwal2011}, namely that $\eta^\perp / \eta^\parallel < 0.05$ everywhere during the last 60 efolds.}\cite{Agarwal2011}. We have
\begin{align}
\eta^\parallel &\equiv \frac{\mathcal{D}_t \dot{\phi}^I G_{IJ} \dot{\phi}^J}{H|\dot{\phi}|^2} \approx \epsilon - \frac{\nabla_\gamma^{(2)} V}{V}, \label{SRetapar} \\
\eta^\perp &\equiv \frac{|(\mathcal{D}_t \dot{\phi})^\perp |}{H|\dot{\phi}|},
\end{align}
where $\perp^{IJ} = G^{IJ} - \dot{\phi}^I \dot{\phi}^J / |\dot{\phi}|^2$ is the projection onto the orthogonal (`entropic') subspace and $\nabla_\gamma^{(2)} = \frac{\dot{\phi}^I}{|\dot{\phi}|} \frac{\dot{\phi}^J}{|\dot{\phi}|} \partial_I \partial_J$ is the covariant second derivative. The approximation in \eqref{SRetapar} holds in the slow roll approximation which is again the immediate generalisation of the single-field slow roll result.

\subsection{Amplitude of Primordial Perturbations and Cosmological Constant}
\label{amppert}

\begin{figure}[h!]
\centering
\includegraphics[width=5in]{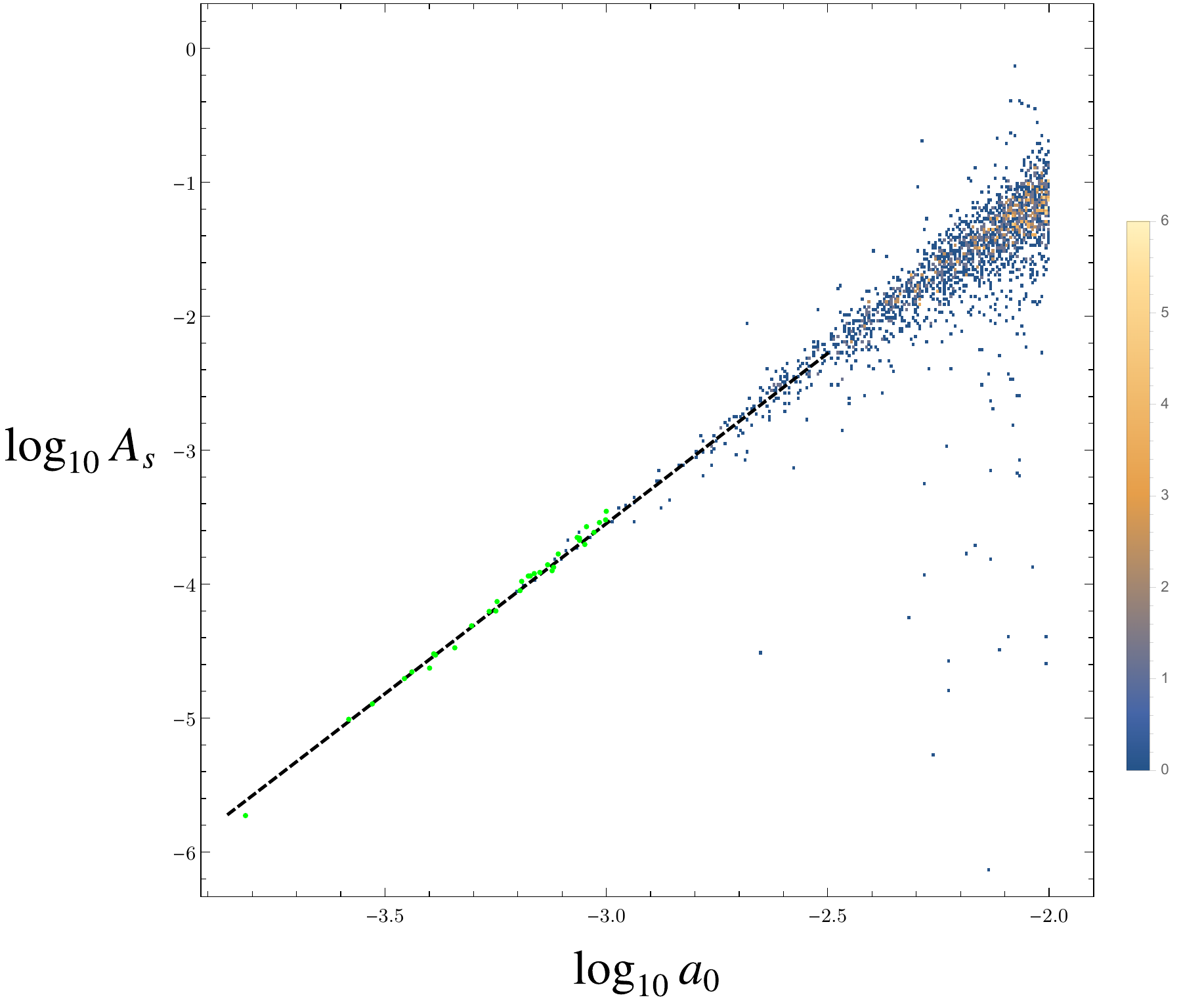}
\caption{Statistics in the $(\log_{10} a_0, \log_{10} A_s)$-plane for realisations with a regime of EI. One sees that eternal inflation patches in the D-brane landscape are concentrated along a narrow line for $\log_{10}a_0 \leq -2$. The NBWF strongly favours small $A_s$ and yields, combined with the statistics, a sharp prediction $A_s = A_s^{\tiny \mbox{obs}}$. The stringency of the EI condition for small $a_0$ means our ensemble of 2106 realisations does not contain instances with the observed scalar amplitude $A_s^{\tiny \mbox{obs}}$. However we have every reason to believe such patches do exist. Green data points originate from a separate scan at small values of $a_0$ and $Q$.}
\label{Asvsa0EI}
\end{figure}

We first discuss the NBWF predictions \eqref{bayes} for the amplitude $A_s$ of the scalar perturbations, taking into account the selections discussed in Section \ref{NBWF}. The no-boundary weighting \eqref{buapprox} strongly favours eternal inflation patches with low values of $a_0$. Fig. \ref{Asvsa0EI} shows this amounts to a strong bias towards low values of $A_s$. In particular, Fig. \ref{Asvsa0EI} reveals that in this regime the eternal inflation patches in the D-brane landscape are concentrated along the following line in the $(\log_{10} a_0, \log_{10} A_s)$-plane,
\be
\log_{10} A_s \sim \frac{5}{2} \log_{10} a_0 + 4
\label{fitAs}
\ee
where both coefficients were determined on the basis of our simulations. If we substitute this in the no-boundary weighting \eqref{buapprox} then the TD probabilities \eqref{bayes} for $(\Dell,A_s)$, marginalised over the remaining perturbation observables, become
\be
\label{ampl}
P(\Dell, A_s| \Dobsaoi) \propto \frac{1}{A_s} A_s^{4/5} \exp \left[\frac{3\pi \cdot 10^{32/5}}{2 T_3 A_s^{8/5}}\right] \theta (A_s - A_s^c)
\ee
where the first factor arises from the logarithmic weighting on $a_0$ employed when we scan the landscape, and the second factor is the statistical distribution of landscape patches. The latter can be derived from Fig. \ref{NEIvsNe}. This shows that patches of EI become exceedingly rare in the landscape for small values of $a_0$. However the no-boundary weighting towards small $A_s$ is much stronger, leading to the prediction that the joint TD probabilities are sharply peaked around the observed value $\log_{10} A_s^c = \log_{10} A_s^{\tiny \mbox{obs}} \approx -8.67$. 

The no-boundary weighting towards small $A_s$ also implies that the predicted dark energy density agrees well with the observed value (cf. Fig. \ref{fraction}). This is in contrast with the prediction of $\Lambda$ based on a uniform prior as discussed in \cite{Hartle2013}.

We should mention that all 2106 patches of EI we identified in our simulations have $A_s > A_s^{\tiny \mbox{obs}}$. Technically speaking the above prediction is based on the extrapolation towards smaller $a_0$ of the curve \eqref{fitAs} that provides an excellent fit to the results of our simulations in the small $a_0$ regime. This indicates that the observed amplitude should be reached at around $\log_{10} a_0 \approx -5$. The fact that our simulations do not contain a realisation of EI with $a_0 \sim 10^{-5}$ is natural on statistical grounds. We have absolutely no reason to expect such EI patches don't exist.

From \eqref{ampl} we see that the scanning measure on $A_s$ (or $a_0$) has a negligible influence on the predictions following from the NBWF. Our choice of numerical bounds $a_0 \in [10^{-5},10^{-2}]$ also clearly has no effect. The same can be said for $Q$, as long as it's allowed to be sufficiently small. Also, including additional contributions $V_0$ to the inflationary vacuum energy will not alter these predictions: the line of data points would simply get shifted upwards or downwards depending on the sign of $V_0$ (see also Section \ref{conclusion}).

Finally we note that \eqref{fitAs} roughly coincides with the maximum value of $\log_{10} A_s$, evaluated 60 efolds before the end of inflation, as a function of $a_0$, in the entire inflationary ensemble. This can be understood from the intuitive picture we have sketched that the EI condition selects the flattest inflection points in the landscape, with the smallest possible $V'$ given $V$. The small slope on the scales associated with EI generally persists throughout the slow roll phase because D-brane inflation is a small field model.

\subsection{Spectral Properties of Primordial Perturbations}
\label{spectralprop}

\begin{figure}[h!]
\centering
\includegraphics[width=4.25in]{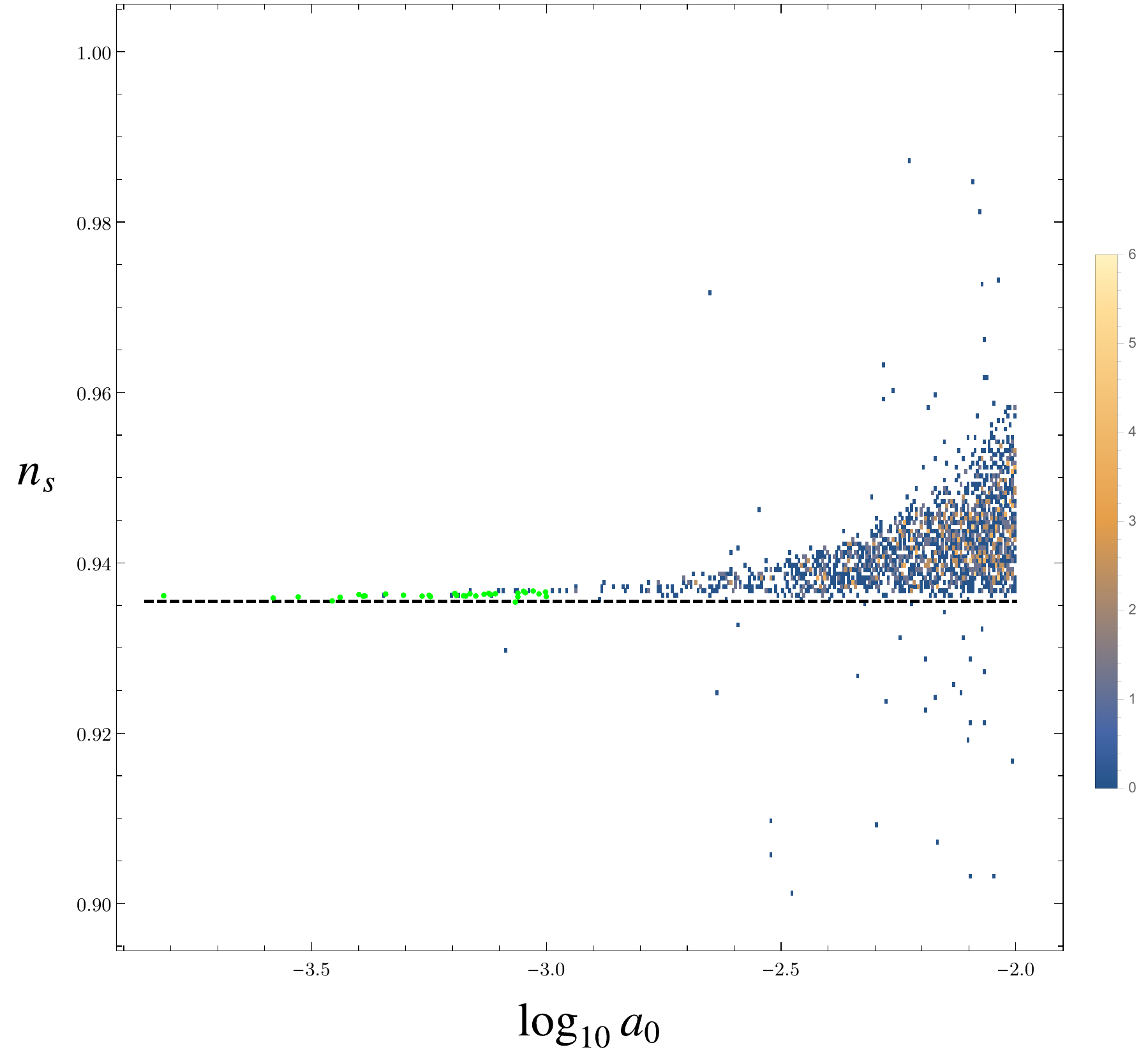}
\caption{Statistical distribution of the scalar tilt $n_s$ in the $(\log_{10} a_0, n_s)$-plane over EI patches in the D-brane landscape. The bias of the NBWF towards small $a_0$ yields a sharp prediction $n_s \approx .936$. This agrees with the value of $n_s$ in the $N_e \rightarrow \infty$ limit of inflection point inflation, see \eqref{nsapprox}. Green data points originate from a separate scan at small values of $a_0$ and $Q$.}
\label{nsvsa0EI}
\end{figure}

\begin{figure}[h!]
\centering
\includegraphics[width=4.25in]{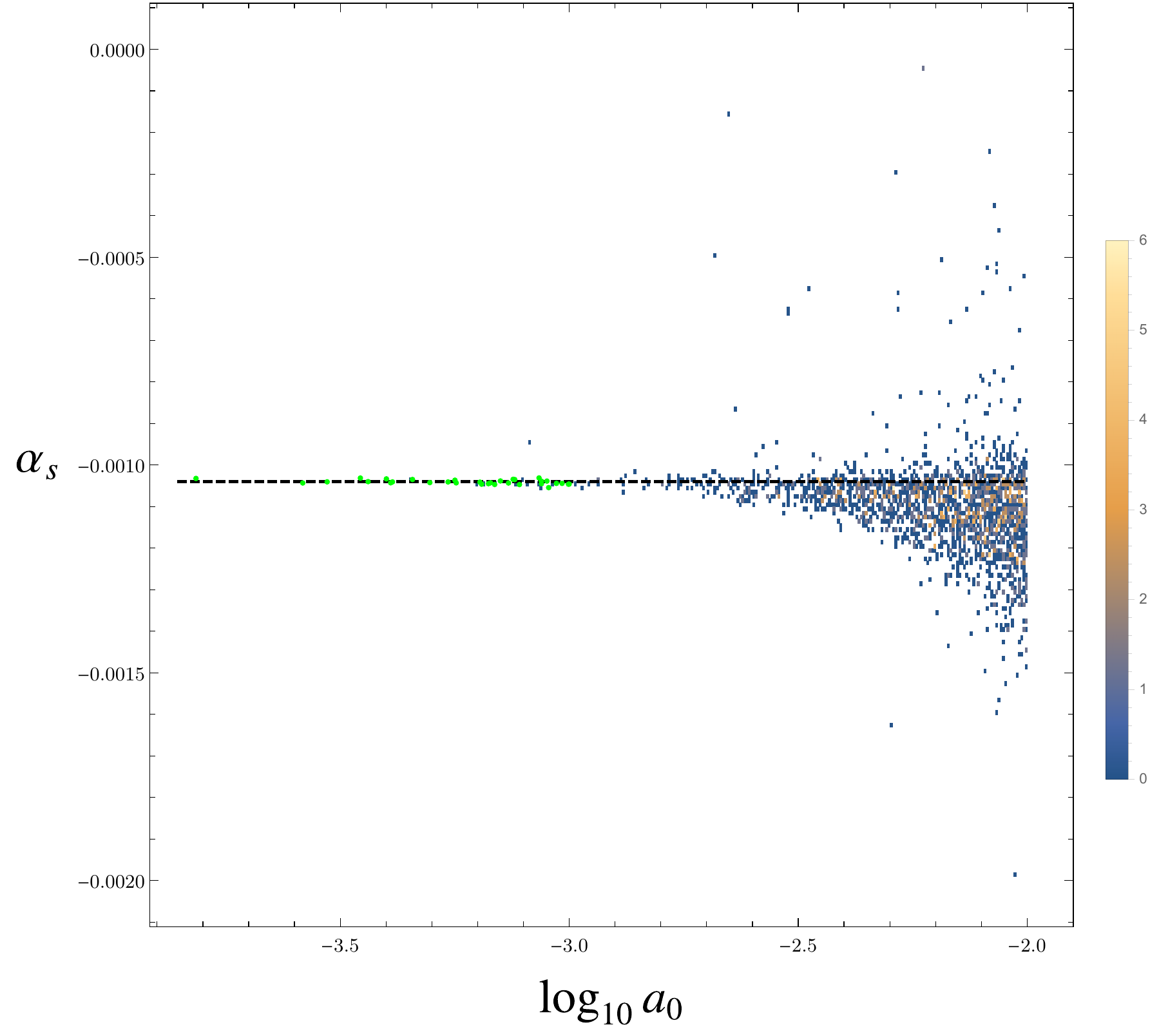}
\caption{Statistical distribution of the running $\alpha_s$ of the scalar tilt  in the $(\log_{10} a_0, \alpha_s)$-plane over EI patches in the D-brane landscape. The bias of the NBWF towards small $a_0$ yields a sharp prediction $\alpha_s \approx -0.00103$. This agrees with the value of $\alpha_s$ in the $N_e \rightarrow \infty$ limit of inflection point inflation, see \eqref{alphasapprox}. Green data points originate from a separate scan at small values of $a_0$ and $Q$.}
\label{alphasvsa0EI}
\end{figure}

\begin{figure}[h!]
\centering
\includegraphics[width=4.25in]{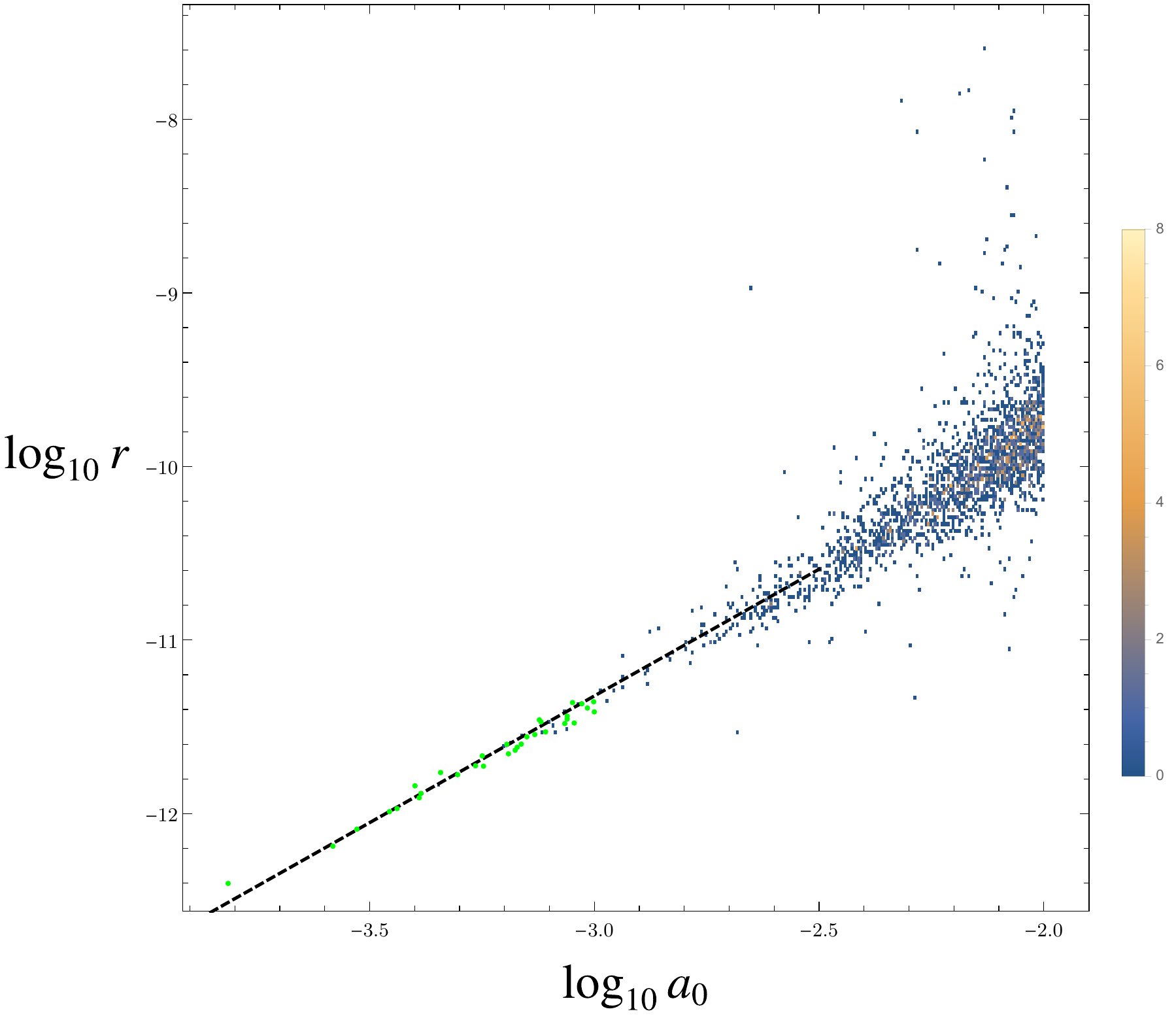}
\caption{Statistical distribution of the tensor-to-scalar ratio $r$ in the $(\log_{10} a_0, \log_{10} r)$-plane over EI patches in the D-brane landscape. The bias of the NBWF towards small $a_0$ yields the prediction $ r \in [-14.85,-13.68]$ at $1 \sigma$.
Green data points originate from a separate scan at small values of $a_0$ and $Q$.}
\label{rvsa0EI}
\end{figure}

We now turn to the remaining observables $n_s, \alpha_s$ and $r$ in $\cO$ associated with the spectral properties of CMB perturbations. The NBWF prior \eqref{bu} selects EI patches with $A_s = A_s^{\tiny \mbox{obs}}$, but it is flat over $n_s, \alpha_s$. However it turns out that the landscape statistics in the restricted subset of EI patches with $A_s = A_s^{\tiny \mbox{obs}}$ leads to definite predictions for $n_s, \alpha_s$. This is evident from our simulations. The data points for these two observables in EI patches are given in Figs. \ref{nsvsa0EI} and \ref{alphasvsa0EI} respectively.

First, Fig. \ref{nsvsa0EI} shows that the no-boundary measure in the D-brane landscape predicts $n_s \approx .936$. This agrees with the large $N_e $ limit of an inflection point model, as we discuss below (see also \cite{Baumann2007}), up to a correction second order in $\eta^{\parallel}$. Indeed, instances of EI with small $a_0$ have many efolds of classical inflation, which are expected to be effectively single-field and inflating along an inflection point characteristic of D-brane inflation. The same reasoning holds for $\alpha_s$, for which the no-boundary measure predicts $\alpha_s \approx -.00103$ (cf. Fig. \ref{alphasvsa0EI}). The $N_e \rightarrow \infty$ limit of an inflection point model yields $\alpha_s \approx -.00104$.

We now turn to the tensor-to-scalar ratio $r$. Here the situation is analogous to that of $A_s$ as can be seen from Fig. \ref{rvsa0EI}, which shows that the EI data points in the small $a_0$ regime are concentrated along a narrow band in the $(\log_{10} a_0, \log_{10} r)$-plane. Not surprisingly this turns out to be a rough lower bound on $r \sim (V'/V)^2$, for a given $a_0$, in the entire $\{ N_e > 60 \}$ ensemble. The extrapolation of the curve that fits the data for $r(a_0)$ to $\log_{10} a_0 \approx -5$ yields the predicted value in the NBWF. Taking the statistical uncertainties (due to a lack of realisations in this regime) into consideration we find $r \in [-14.55,-13.74]$ at the 68\% confidence level.

Finally, we remark that signatures of non-Gaussianity as encoded by the parameters $f_{NL}$ and $t_{NL}$ are expected to be unobservably small in the EI ensemble. This follows from the estimates given in \cite{Mcallister2012}, from which the very conservative estimates $f_{NL} < 10^{-4}, t_{NL} < 10^{-7}$ can be deduced for the EI ensemble. This is orders of magnitude smaller than the current error bars on the observed values \cite{Ade2015}. Hence the NBWF predicts no observable non-Gaussianity.

\section{Inflection Point Model}
\label{inflectionptmodel} 

It is well known that a simple model of inflection point inflation captures a good part of the phenomenology of the classical cosmology of D-brane inflation (see e.g. \cite{Baumann2007}). In this Section we show that a single-field model of inflection point inflation describes also most of the phenomenology of D-brane inflation in quantum cosmology, in particular the predictions associated with eternal inflation patches. We consider the following cubic potential \cite{Baumann2007},
\begin{equation}
\label{model}
V(\phi) = V_0 + \lambda_1 \phi + \frac{\lambda_3}{3!}\phi^3,
\end{equation}
with $V_0, \lambda_1, \lambda_3 > 0$, $V_0 \ll 1$ and $V_0 \gg \lambda_1 \phi + \frac{\lambda_3}{3!}\phi^3$ during inflation. The slow roll parameters are
\be
\epsilon \approx \frac{1}{2} \left( \frac{\lambda_1 + \frac{\lambda_3}{2} \phi^2}{V_0} \right)^2, \qquad
\eta = \frac{\lambda_3}{V_0}\phi.
\ee
At the inflection point the conditions for inflation must hold, so $\epsilon \ll 1$ at $\phi = 0$ and hence $V_0 \gg \lambda_1$. Inflation ends\footnote{With $V_0 \gg \lambda_1$ and under the assumption of small field inflation, $|\eta|$ becomes unity before $|\epsilon|$ does.} when $ \eta = -1,$ or $\phi_e = - V_0 / \lambda_3$. Thus eq. \eqref{model} is a small field inflation model if
\begin{equation} \label{smallfield}
V_0 \ll \lambda_3.
\end{equation}
The total number of efolds of the \textit{background} cosmology that probes the entire inflationary patch (i.e. in which the scalar starts at $\eta \geq 1$) is given by
\begin{equation} \label{bkgr-efolds1}
N_{\tiny \mbox{tot}} \approx 4 \sqrt{\frac{V_0^2}{2 \lambda_1 \lambda_3}} \arctan \left( \sqrt{\frac{V_0^2}{2\lambda_1 \lambda_3}} \right),
\end{equation}
which we see will be large (Fig. \ref{Nevsa0EI}) when
\begin{equation} \label{bound2}
\frac{V_0}{\sqrt{2 \lambda_1 \lambda_3}} \gg 1
\end{equation}
such that
\begin{equation} \label{bkgr-efolds2}
N_{\tiny \mbox{tot}} \approx \pi \sqrt{\frac{2 V_0^2}{\lambda_1 \lambda_3}} \gg 1.
\end{equation}

In order for the EI condition to hold somewhere along the trajectory - and hence for \eqref{model} to model an EI patch in the D-brane landscape - it suffices to require EI occurs at the inflection point. In the slow roll approximation the EI condition is $V_0 > 24 \pi^2 \epsilon\big|_{\phi = 0}$, which gives 
\begin{equation} \label{bound1}
\lambda_1^2 < \frac{1}{12 \pi^2} V_0^3 \equiv \bar{\lambda}_1^2
\end{equation}
where we have defined the threshold value $\bar{\lambda}_1$. This is in line with our experience in the D-brane landscape where we find that EI patches become increasingly rare as $V_0$ decreases. The regime of eternal inflation of a typical EI realisation occurs only in a small neighbourhood $\Delta \phi \sim 2 V_0 / \lambda_3$ of the inflection point. On general grounds we expect $\lambda_1 \approx \bar{\lambda}_1$ in a typical realisation. Combining \eqref{smallfield} with \eqref{bound2} and \eqref{bound1} implies $V_0 \ll \lambda_3 \ll V_0^{1/2}$. We take
\begin{equation}
\label{fit}
\lambda_3 = V_0^\alpha
\end{equation}
where we consider $\frac{1}{2} < \alpha < 1$ to be a fit parameter when comparing \eqref{fit} with the results of our simulations. Notice that the field space extent of the EI patch is $2 V_0 / \lambda_3 \gg \sqrt{V_0}$. During eternal inflation quantum fluctuations of the scalar field are of order $H \sim \sqrt{V}$, so the EI patch in small field inflection point inflation with $N_{\tiny \mbox{tot}} \gg 1$ is guaranteed to be sufficiently large to ensure that inflation is actually eternal. Using \eqref{bkgr-efolds2}, at $N_* $ efolds before the end of inflation we have
\begin{equation} \label{phistar}
\phi_* \approx \sqrt{\frac{2\lambda_1}{\lambda_3}} \tan \left[ \frac{\pi N_*}{N_{\tiny \mbox{tot}}} - \arctan \left( \frac{N_{\tiny \mbox{tot}}}{2\pi} \right) \right] \approx \frac{-2}{N_*} V_0^{1-\alpha}.
\end{equation}
where the last equality holds if $N_{\tiny \mbox{tot}} \gg N_*$ or equivalently $V_0 \ll N_*^{\frac{4}{1-2\alpha}}$.
It follows that $V(\phi_*) \approx V_0$ and, keeping only the leading terms, that
\begin{align}
A_s &\approx \frac{N_*^4}{48 \pi^2} V_0^{2\alpha - 1}, \label{Asapprox} \\
n_s &\approx 1+ 2 \eta_* \approx 1 - \frac{4}{2 + N_*} + \frac{4 \pi^2}{3} \frac{N_*}{N_{\tiny \mbox{tot}}^2} + \mathcal{O}\left( N_{\tiny \mbox{tot}}^{-4} \right), \label{nsapprox} \\
\alpha_s &= - \frac{\mathrm{d}n_s}{\mathrm{d}N_*} \approx \frac{-4}{(2+N_*)^2}, \label{alphasapprox} \\
r &= 16 \epsilon_* \approx \frac{32}{N_*^4} V_0^{2(1-\alpha)}. \label{rapprox}
\end{align}

These values of $n_s$ and $\alpha_s$ for $N_* \sim 60$ agree very well the results of our simulations  summarised in Fig. \ref{nsvsa0EI} and Fig. \ref{alphasvsa0EI}. For \eqref{Asapprox} and \eqref{rapprox}, we can use $V_0 \approx D_0 = 2 T_3 a_0^4$ to rewrite
\begin{align}
\log_{10} A_s &\sim 4(2\alpha - 1) \log_{10} a_0, \label{Asapprox2} \\
\log_{10} r &\sim 8(1-\alpha) \log_{10} a_0. \label{rapprox2}
\end{align}
where we concentrate on the slope. Best fit lines to the small $a_0$ data in the ($\log_{10} a_0, \log_{10} A_s$) and ($\log_{10} a_0, \log_{10} r$)-planes (Figs. \ref{Asvsa0EI}, \ref{rvsa0EI}) give slopes 2.54011 and 1.45987 respectively. One can verify that $\alpha \approx 0.817513$ according to \eqref{rapprox2}, and $\alpha \approx 0.817517$ according to \eqref{Asapprox2} which is nicely consistent. \\

Eq. \eqref{Asapprox} together with the fact that $V \approx V_0$ at the EI threshold also yields the approximate no-boundary weighting $\eqref{ampl}$ we used earlier in the D-brane landscape.

\section{Alternative Measures and Observables}
\label{othermeasures}

The sharp predictions based on the NBWF contrast with the broad distributions predicted on the basis of the usual ad hoc uniform measure in the D-brane landscape, augmented with the assumption that $N_e > 60$ (and without taking into account our observational situation). This suggests different theories of the universe's quantum state can to some extent be observationally distinguished and thereby tested. On the other hand we have no evidence that the NBWF is the unique state exhibiting the correlation that we discussed. Any specific quantum state that selects inflation is likely to predict at least some of these correlations. This is because the top-down selection of patches of eternal inflation does not rely on the specific input supplied by the NBWF. It plausibly applies more generally. This suffices to obtain precise predictions for the joint probabilities of the set of local data $\Dell$ and the spectral properties of CMB perturbations: the probability distribution for $\Dell$ implies $A_s$ lies within a rather limited range, which in the D-brane landscape yields sharp predictions for the remaining CMB-related observables (cf. e.g. Fig. \ref{nsvsa0EI}). In short, the quantum state selects EI and the structure of the landscape potential does the rest.

However this simple scheme does not apply to our predictions of the amplitude $A_s$. Here we do rely on the NBWF bias towards small values of $A_s$ within the EI ensemble. Adopting a uniform measure on the EI ensemble may not significantly change this but if one were to adopt the tunnelling state \cite{Vilenkin1987}, which in the EI landscape strongly favours universes with large $A_s$, then one would predict $\log_{10} A_s \sim -5$ much larger than its observed value.

One might wonder what would happen to our predictions if some of the observables had been left out \cite{Hartle2013}.
Suppose we had calculated the joint probability describing a correlation between $\Lambda$, $A_s$, and a more restricted set of local data $\Dell$ that do not include a specification of the age. To examine whether the NBWF predicts a correlation between the observed values of $\Lambda$, $A_s$ and the observation that $\Dell$ exists {\it some time} one must evaluate the joint probability \eqref{bayes}, where $P(\Dell|\Lambda, A_s)$ is again of the form \eqref{fractgal} but now with the fraction $f$ replaced by its asymptotic value and without the step function in time. The lower bound on $A_s$ is no longer set by the age $t_0$ as in Fig. \ref{fraction}, but by the requirement that there be sufficient radiative cooling of collapsed halos for stars to form \cite{Tegmark2006}.

The NBWF favours the smallest $A_s$ in the region selected by $P(\Dell|\Lambda, A_s)$ which is $A_s \sim 10^{-12}$. In those universes galaxies typically form at a time later than $t_0$ from the collapse of perturbations which have an initial amplitude $A_s$ smaller than the observed value. The NBWF prior assigns much lower probability to a correlation between the observed values of $\Lambda$, $A_s$ and the existence of $\Dell$ some time, than it does to the correlation between $\Lambda$, $A_s$ and $\Dell$ that includes a specification of the age $t_0$. Does this mean the theory has failed in its predictive power? No, because in quantum cosmology one expects only some correlations to be predicted with significant probabilities. We need to search to find these, and the comparison of these two correlations is an example of this search. Data that we have should not be left out unless it can be demonstrated they do not affect the result. Indeed, if we had neglected the formation of stars in addition to the age, $A_s$ would have been pushed down to ${\cal O}(10^{-20})$.

On the other hand, in most of the literature on D-brane inflation it is simply {\it assumed} that the amplitude $A_s$ takes its observed value. In quantum cosmology this is done by including our observations of $A_s$ in the data $\Dobs$ that are conditioned on in \eqref{res}. It is immediately clear this would yield extremely sharp predictions for the remaining observables ${\cal O}$ in \eqref{bayes}. This provides perhaps the cleanest illustration of the contrast between the sharp NBWF predictions in D-brane inflation and the broad distributions implied by a uniform measure on the entire $\{ N_e > 60 \}$ ensemble.

\section{Conclusion}
\label{conclusion}

We have shown that the primordial perturbation patterns in universes emerging from low eternal inflation patches in the D-brane landscape share very specific spectral properties. These include a scalar tilt $n_s \approx .936$, a running of the scalar tilt $\alpha_s \approx -.00103$, a tensor-to-scalar ratio $\log_{10} r < -13$, undetectably small non-Gaussianity and no observable spatial curvature.

The NBWF applied to D-brane inflation predicts that our universe has emerged from one of those regions of eternal inflation. Consequently the top-down probabilities for CMB-related observables in the no-boundary state in D-brane inflation are sharply peaked around the above values. By contrast either the D-brane landscape or the NBWF alone yields broadly distributed probabilities for those observables and hence no precise predictions. This explicitly demonstrates that both the structure of the landscape - the so-called `statistics' - and the quantum state are essential for cosmology to be predictive.

It hasn't escaped our notice that the sharp prediction for the spectral tilt in D-brane inflation lies outside the range allowed by observations \cite{Planck_inflation}. It is an important question whether structure of the D-brane landscape we have not taken into account can alleviate this tension. Fig. \ref{nsvsa0EI} shows there do exist EI histories for larger values of $a_0$ with values of $n_s$ that lie within the range allowed by observations. These histories have far fewer efolds after the exit from EI and an amplitude $A_s$ that is much larger than its observed value. At first sight it appears one can adjust $A_s$ independently by including contributions $V_0$ to the inflationary potential from other regions of the Calabi-Yau, along the lines of \cite{Mcallister2012}. If those contributions lower the potential then the observed $A_s$ will correspond to a larger value of $a_0$. However it appears that the observationally viable histories with larger values of $n_s$ no longer satisfy the EI criterion when $V_0$ lowers the potential significantly. The EI histories that remain all have many efolds of slow roll after the EI exit and $n_s \approx .936$. This is also borne out by the phenomenological inflection point model.

We are therefore led to the conclusion that observations rule out D-brane inflation for a wide range of measures, including the NBWF, that select EI patches in the landscape. However this disagreement with observation is not our main result. After all this is but a particular corner of the string landscape. Rather we have an important conceptual result, namely that precision cosmology probes an intricate combination of the statistical properties of the string landscape and the universe's quantum state and therefore can be employed to test both parts of the theoretical framework.

\vskip 1cm

\noindent{\bf Acknowledgements:} We thank Tolya Dymarsky, Jonathan Frazer, James Hartle, Rob Speare, Thomas Bachlechner and Matt Kleban for useful discussions. We thank the KITP and the Physics Department at UCSB for their hospitality. The work of TH is supported in part by the Belgian National Science Foundation (FWO) grant G.001.12 Odysseus and by the European Research Council grant no. ERC-2013-CoG 616732 HoloQosmos.

\bibliographystyle{utphys}
\bibliography{DbraneNBWF}

\end{document}